\documentclass[twocolumn,preprintnumbers,amsmath,amssymb,subfig,aps,prb,nobibnotes,showkeys,doi]{revtex4-1}
\usepackage{graphicx}% Include figure files
\usepackage{hhline}
\usepackage{xcolor}
\usepackage{natbib}
\setcitestyle{numbers,square}

\begin{document}

\title{Origin of the classical magnetization discontinuities of the dodecahedron}

\author{N. P. Konstantinidis}
\affiliation{Mathematics and Science Department, American University in Bulgaria, Sv. Buchvarova Str. 8, Blagoevgrad 2700, Bulgaria}

\date{\today}

\begin{abstract}
The classical antiferromagnetic Heisenberg model on the dodecahedron has been shown to have three magnetization discontinuities in an external field. Here it is shown that the highest-field discontinuity can be directly traced back to the strong magnetization jump leading to saturation at the Ising limit, which originates from the magnetic response of an isolated pentagon and the frustrated connectivity of the dodecahedron. This discontinuity survives up to the $XY$ limit and disappears shortly before the ferromagnetic Ising interaction fully polarizes the spins. The two lower-field jumps of the model result from the competition of discontinuities that emerge from the magnetization plateau surviving away from the Ising limit.
\end{abstract}


%\pacs{75.10.Hk Classical Spin Models, 75.50.Ee Antiferromagnetics, 75.50.Xx Molecular Magnets}

\maketitle

\section{Introduction}
\label{sec:introduction}

\begin{figure}[h]
\begin{center}
\includegraphics[width=2.3in,height=3.7in,angle=-90]{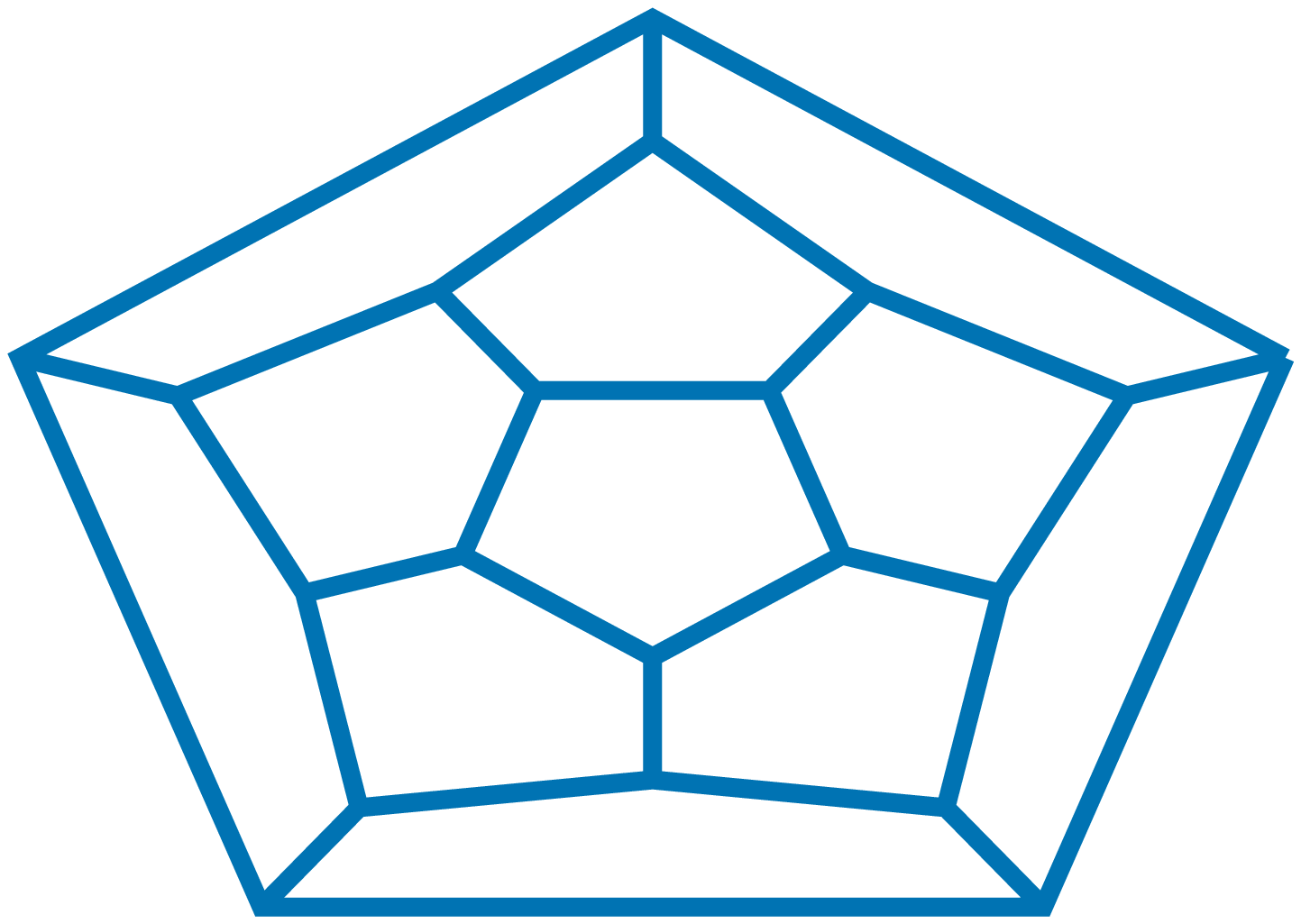}
\end{center}
\caption{Planar projection of the dodecahedron.
%(~/basic/classical/fullerenes)
}
\label{fig:dodecahedron}
\end{figure}

The dodecahedron (Fig. \ref{fig:dodecahedron}) is a Platonic solid \cite{Plato} with 20 vertices, which are all geometrically equivalent. It belongs to the class of Goldberg polyhedra \cite{Goldberg37} and consists of twelve pentagons, and is the smallest fullerene in the form of C$_{20}$ \cite{Fowler95,Prinzbach00,Wang01,Iqbal03,Qin17}. All of its edges are symmetrically equivalent. It transforms according to the icosahedral-$I_h$ point group, the largest point group with 120 symmetry operations \cite{Altmann94}.

The antiferromagnetic Heisenberg model (AHM) describes interactions between localized spins existing in three spin-space dimensions \cite{Auerbach98,Fazekas99}. The spins are mounted on the vertices of lattices or molecules. A case of special interest is when antiferromagnetic interactions between nearest-neighbor classical spins do not result in them being antiparallel in the ground state of the AHM. This is due to competing interactions and is known as frustration \cite{Lhuillier01,Misguich03,Ramirez05,Schnack10}.
%,Florek19,Schmidt20,Schmidt20-1,Schmidt22}.
The dodecahedron is such an example, as the pentagons from which it is made are frustrated. The classical AHM on an isolated pentagon has a ground state with all spins lying in the same plane and an energy per bond equal to $-\frac{\sqrt{5}+1}{4}$ \cite{Schmidt03}. In the case of the dodecahedron the ground state is three-dimensional with an energy per bond equal to $-\frac{\sqrt{5}}{3}$ \cite{Coffey92}, higher than the one of an isolated pentagon, showing that the dodecahedron's connectivity introduces further frustration from the one at the isolated pentagon level.

Another important consequence of frustration for the classical AHM on the dodecahedron is the discontinuous ground-state magnetization in an external field, even though the AHM lacks anisotropy in spin space. In total there are three discontinuities \cite{NPK07}, and the magnetization response is also discontinuous at the full quantum limit \cite{NPK05,NPK23-1}.
%These jumps are not due to any magnetic anisotropy, but rather the frustrated connectivity of the dodecahedron.
Implications for thermodynamic properties of the dodecahedron have also been examined \cite{Strecka15,Karlova16,Karlova16-1,Karlova16-2}. Other fullerenes of icosahedral and different symmetries have also been found to exhibit rich magnetic behavior \cite{Coffey92,NPK09,NPK17,NPK18}. The AHM on the icosahedron, the dual of the dodecahedron, has a ground-state magnetization discontinuity at the classical level \cite{Schroeder05}, which is directly related to its connectivity \cite{NPK15}. Classical magnetization jumps have also been found for the 600-cell, a four-dimensional analog of a Platonic solid, for the AHM and XY models \cite{NPK23-2}. The pentakis dodecahedron, the dual of the truncated icosahedron, presents discontinuities at both the classical and quantum levels \cite{NPK21}. Extended systems are also associated with quantum magnetization discontinuities in the ground state \cite{Schulenburg02,Richter04,Schnack06,Nakano13,Nakano14,Nakano14-1,Furuchi21,Furuchi23}. %Finally, frustration has been linked with the search for a spin liquid state \cite{Savary16,Zhou17,Knolle19,Ramirez24}.

The origin of the quantum magnetization discontinuity in the isotropic antiferromagnetic Heisenberg limit has been traced to a strong discontinuity in the Ising limit, which survives for infinitesimal fluctuations in the $xy$ plane \cite{NPK23-1}. This discontinuity has been shown to persist up at least to the isotropic limit by allowing the fluctuations in the $xy$ plane to become progressively stronger. This is true for individual spin quantum numbers $s=\frac{1}{2}$ and 1, and also has important consequences for the magnetic response of bigger fullerenes that share the dodecahedron's symmetry.

In this paper the same procedure is followed but for spins which are classical, in order to trace the origin of the three discontinuities of the AHM in the Ising limit. As the interactions away from the Ising axis are added the Ising-limit jump splits into two, and the resulting higher-field discontinuity is present not only at the Heisenberg but also at the $XY$ limit. It further survives the ferromagnetic Ising interactions quite close to the ferromagnetic limit, where they fully polarize the spins. It is shown that the Ising-limit discontinuity exists in the magnetic response of an isolated pentagon, the basic constituent of the dodecahedron, but disappears just before the Heisenberg limit, pointing to the importance of the connectivity of the dodecahedron for the survival of the higher-field jump almost up to the ferromagnetic limit. The lower-field jump together with other discontinuities that emerge for sufficiently strong planar interactions generate a rich discontinuous magnetization and susceptibility response close to the isotropic limit for lower fields. This eventually results in the two lower-field magnetization discontinuities of the AHM.

The results of this paper show that one of the classical AHM magnetization discontinuities can be directly traced back to the Ising-limit discontinuity, with the latter a direct consequence of the magnetic response of an isolated pentagon and the frustrated connectivity of the dodecahedron. This is similar to what was shown for the quantum discontinuity for $s \leq 1$ \cite{NPK23-1}, however the lack of a jump for finite quantum numbers $s>1$ distinguishes between the discontinuity mechanism in the classical and quantum case. It is concluded that at the full quantum and the classical limit the AHM on the dodecahedron is associated with discontinuous magnetization response, even though there is no magnetic anisotropy, due to its frustrated connectivity.

The plan of this paper is as follows: Sec. \ref{sec:model} introduces the anisotropic Heisenberg model, Sec. \ref{sec:groundstateinzeromagneticfield} presents the ground state in the absence of a field, and Sec. \ref{sec:groundstatemagnetizationinanexternalfield} the magnetic response in an external field. Sec. \ref{sec:conclusions} discusses the conclusions.

\section{Model}
\label{sec:model}
The Hamiltonian of the anisotropic Heisenberg model is

\begin{eqnarray}
H = \sum_{<ij>} [ sin\omega ( s_i^x s_j^x + s_i^y s_j^y ) + cos\omega s_i^z s_j^z ] - h \sum_{i=1}^{N} s_i^z
\label{eqn:model}
\end{eqnarray}

The dodecahedron has $N=20$ vertices, with each one three-fold coordinated. On each vertex $i=1,\dots,N$ a classical spin $\vec{s}_i=s_i^x\hat{x}+s_i^y\hat{y}+s_i^z\hat{z}$ is mounted. The first term in Hamiltonian (\ref{eqn:model}) describes the exchange interactions between the spins. The brackets in $<ij>$ indicate that the interactions are limited to the 30 nearest-neighbor pairs. All the edges are symmetrically equivalent, making all interactions equal. The exchange interactions define the unit of energy and are parametrized as $cos\omega$ along the $z$ axis and $sin\omega$ in the $xy$ plane. The second term in Hamiltonian (\ref{eqn:model}) is the energy due to an external magnetic field of magnitude $h$, taken along the $z$ axis. Here the region $0 \leq \omega \leq \pi$ is considered. $\omega=0$ corresponds to the antiferromagnetic Ising model in a parallel magnetic field, $\omega=\frac{\pi}{4}$ to the AHM in a field, and $\omega=\frac{\pi}{2}$ to the antiferromagnetic $XY$ model in a transverse magnetic field. For $\frac{\pi}{2} < \omega \leq \pi$ the Ising interaction becomes ferromagnetic. The saturation magnetic field $h_{sat}=3cos\omega+\sqrt{5}sin\omega$ (App. \ref{appendix:saturationmagneticfielddodecahedron}). It becomes zero when $\omega=\pi-tan^{-1}\frac{3}{\sqrt{5}}$, where the Ising interaction is strong enough to force the spins to be parallel even in the absence of a field. The spins $\vec{s}_i$ are unit vectors determined by a polar $\theta_i$ and an azimuthal $\phi_i$ angle.

The lowest-energy configuration of Hamiltonian (\ref{eqn:model}) results from the competition for minimization between the exchange and the magnetic energy, with frustration playing an essential role. Minimization of the Hamiltonian gives the lowest-energy spin configuration as a function of $\omega$ and $h$ \cite{Coffey92,NPK07,NPK16-1}. For each such pair random values are assigned to the polar and azimuthal angles and each one of them is moved opposite the direction of its gradient, until the energy minimum is reached. This procedure is repeated for different initial configurations to ensure that the global energy minimum is found. Each magnetization discontinuity is assigned a number and a color as they appear with increasing $\omega$.

\section{Ground State in Zero Magnetic Field}
\label{sec:groundstateinzeromagneticfield}

Fig. \ref{fig:dodecahedronzerofieldenergy} plots the ground-state energy per bond of Hamiltonian (\ref{eqn:model}) in the absence of a magnetic field as a function of $\omega$. At the Ising limit $\omega=0$ the average energy per bond equals $-\frac{3}{5}$, and the total spin $M$ can be either 0 or 4 \cite{NPK23,NPK23-1}. Introduction of the $xy$-plane interaction for $\omega > 0$ does not change the Ising-type lowest-energy configuration for small $\omega$. This is shown in Fig. \ref{fig:dodecahedronzerofieldenergycontr} that plots the average nearest-neighbor correlation along the $z$ axis and in the $xy$ plane, $\frac{1}{30} \sum_{<ij>} s_i^z s_j^z$ and $\frac{1}{30} \sum_{<ij>} (s_i^x s_j^x + s_i^y s_j^y)$ respectively. The lowest-energy configuration remains the same up to
%$\omega=0.1443415 \pi$
$\omega=0.14434 \pi$, where the planar correlation starts to decrease at the expense of the Ising one, with the spins now forming a three-dimensional ground state. The unique nearest-neighbor correlations $\vec{s}_i \cdot \vec{s}_j$ are shown in Fig. \ref{fig:dodecahedronzerofieldcorr1}. At
%$\omega=0.208076 \pi$
$\omega=0.20808 \pi$ the average ground-state energy per bond becomes maximum and equal to
%-0.5133948
-0.51339. At this value of $\omega$ the Ising and planar correlations are discontinuous.
%The average Ising and planar dot products become equal at $\omega=0.21739630 \pi$, where they are equal to
%-0.10966366.
%-0.3655455.
%The average Ising and planar energies become equal at $\omega=0.222801530 \pi$, where they are equal to -0.257744590.
The ground-state nearest-neighbor correlations, which have been converging going away from the Ising ground state, become equal for the AHM. Another discontinuity of the correlations occurs at
%$\omega=0.27235238\pi$
$\omega=0.27235 \pi$ and leads to a lowest-energy configuration with the spins lying completely in the $xy$ plane. This configuration has six unique nearest-neighbor correlations (Fig. \ref{fig:dodecahedronzerofieldcorr1}) and at the $XY$ limit the ground-state energy per bond achieves a local minimum equal to
%-0.717691168422
-0.71769. The ground state does not change up to
%$\omega=0.69550748 \pi$
$\omega=0.69551 \pi$ where its energy achieves a local maximum equal to
%-0.58652016
-0.58652, the Ising energy starts to decrease at the expense of the energy in the $xy$ plane (Fig. \ref{fig:dodecahedronzerofieldenergycontr}), and the correlations are discontinuous (Fig. \ref{fig:dodecahedronzerofieldcorr2}). At this value of $\omega$ the ground-state magnetization becomes finite for the first time away from the Ising ground state of small $\omega$, acquiring a finite value $M=13.26504$ via a jump. At $\omega=\pi-tan^{-1}\frac{3}{\sqrt{5}}$ the ground state becomes ferromagnetic along the Ising axis.

\begin{figure}
\includegraphics[width=3.5in,height=2.5in]{dodecahedronzerofieldenergy}
\vspace{0pt}
\caption{Ground-state energy per bond $\frac{E_g}{30}$ of Hamiltonian (\ref{eqn:model}) in zero magnetic field as a function of $\omega$.
%(~/basic/classical/dodecahedronparameter/zerofield)
}
\label{fig:dodecahedronzerofieldenergy}
\end{figure}

\begin{figure}
\includegraphics[width=3.5in,height=2.5in]{dodecahedronzerofieldenergycontr}
\vspace{0pt}
\caption{Average ground-state nearest-neighbor correlation along the Ising axis $\frac{1}{30} \sum_{<ij>} s_i^z s_j^z$ (black solid line) and in the $xy$ plane $\frac{1}{30} \sum_{<ij>} (s_i^x s_j^x + s_i^y s_j^y)$ (red dashed line) of Hamiltonian (\ref{eqn:model}) in zero magnetic field as a function of $\omega$.
%(~/basic/classical/dodecahedronparameter/zerofield)
}
\label{fig:dodecahedronzerofieldenergycontr}
\end{figure}

\begin{figure}
\includegraphics[width=3.5in,height=2.5in]{dodecahedronzerofieldcorr1}
\vspace{0pt}
\caption{Unique ground-state nearest-neighbor correlations $\vec{s}_i \cdot \vec{s}_j$ of Hamiltonian (\ref{eqn:model}) in zero magnetic field as a function of $\omega$ for lower $\omega$.
%(~/basic/classical/dodecahedronparameter/zerofield)
}
\label{fig:dodecahedronzerofieldcorr1}
\end{figure}

\begin{figure}
\includegraphics[width=3.5in,height=2.5in]{dodecahedronzerofieldcorr2}
\vspace{0pt}
\caption{Unique ground-state nearest-neighbor correlations $\vec{s}_i \cdot \vec{s}_j$ of Hamiltonian (\ref{eqn:model}) in zero magnetic field as a function of $\omega$ for higher $\omega$.
%(~/basic/classical/dodecahedronparameter/zerofield)
}
\label{fig:dodecahedronzerofieldcorr2}
\end{figure}

\section{Ground-State Magnetization in an External Field}
\label{sec:groundstatemagnetizationinanexternalfield}

\begin{figure}
\includegraphics[width=2.5in,height=3.5in,angle=270]{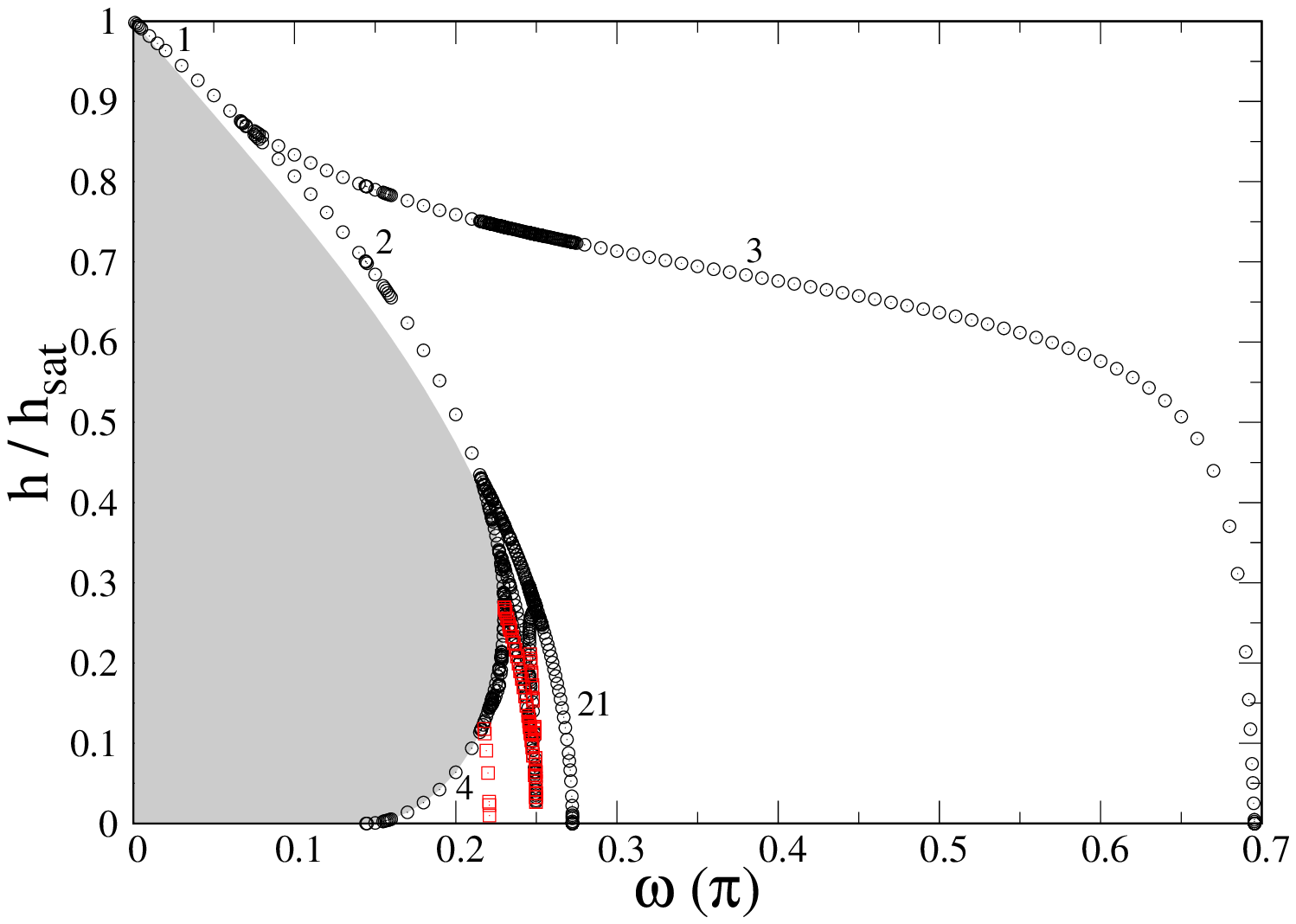}
\vspace{0pt}
\caption{Location of ground-state magnetization (black circles) and susceptibility (red squares) discontinuities and magnetization plateau (gray-shaded area) of Hamiltonian (\ref{eqn:model}) as a function of $\omega$ and the magnetic field $h$ over its saturation value $h_{sat}$. Magnetization discontinuities that are clearly visible are assigned a number as they appear with increasing $\omega$ (see also Fig. \ref{fig:dodecahedrondisc1}).
%(~/basic/classical/dodecahedronparameter)
}
\label{fig:dodecahedrondisc}
\end{figure}

Fig. \ref{fig:dodecahedrondisc} plots the location of the ground-state magnetization and susceptibility discontinuities and the magnetization plateau as a function of $\omega$ and the magnetic field $h$ over its saturation value $h_{sat}$. At the Ising limit $\omega=0$ an infinitesimal field selects the $M=4$ lowest-energy configuration that remains the ground state until saturation, which enters with a magnetization jump $\Delta M=16$ (Fig. \ref{fig:dodecahedronmagnetization}) \cite{NPK23-1}. As the $xy$-plane interaction is switched on for $\omega > 0$ the $M=4$ magnetization plateau persists and the jump survives close to saturation (number 1 in Figs \ref{fig:dodecahedrondisc} and \ref{fig:dodecahedrondiscwidth}), however the inaccessible magnetization range due to the discontinuity is now significantly reduced with respect to the Ising case, as also shown by the magnetization curve for $\omega=0.05 \pi$ in Fig. \ref{fig:dodecahedronmagnetization}. The range of discontinuity 1 is getting wider with $\omega$, something that can also be seen in Fig. \ref{fig:dodecahedrondiscwidth2} that plots the magnitude of all discontinuities as a function of $\omega$. Eventually the high-field discontinuity splits into two at $\omega=0.06766 \pi$ (numbers 2 and 3), with the higher-field jump 3 surviving above the $XY$ limit ($\omega=\frac{\pi}{2}$) up to
%$\omega=0.69550748 \pi$
$\omega=0.69551 \pi$ (Sec. \ref{sec:groundstateinzeromagneticfield}), where the Ising interaction has become ferromagnetic. This demonstrates that the origin of the higher-field discontinuity of the AHM ($\omega=\frac{\pi}{4}$) is the jump appearing at the higher end of the $\omega=0$ plateau, similarly to the quantum case of $s \leq 1$ \cite{NPK23-1}.

\begin{figure}
\includegraphics[width=3.5in,height=2.5in]{dodecahedronmagnetization}
\vspace{0pt}
\caption{Ground-state magnetization $M$ of Hamiltonian (\ref{eqn:model}) as a function of the magnetic field $h$ over its saturation value $h_{sat}$ for different values of $\omega$.
%(~/basic/classical/dodecahedronparameter)
}
\label{fig:dodecahedronmagnetization}
\end{figure}

\begin{figure}
\includegraphics[width=2.5in,height=3.5in,angle=270]{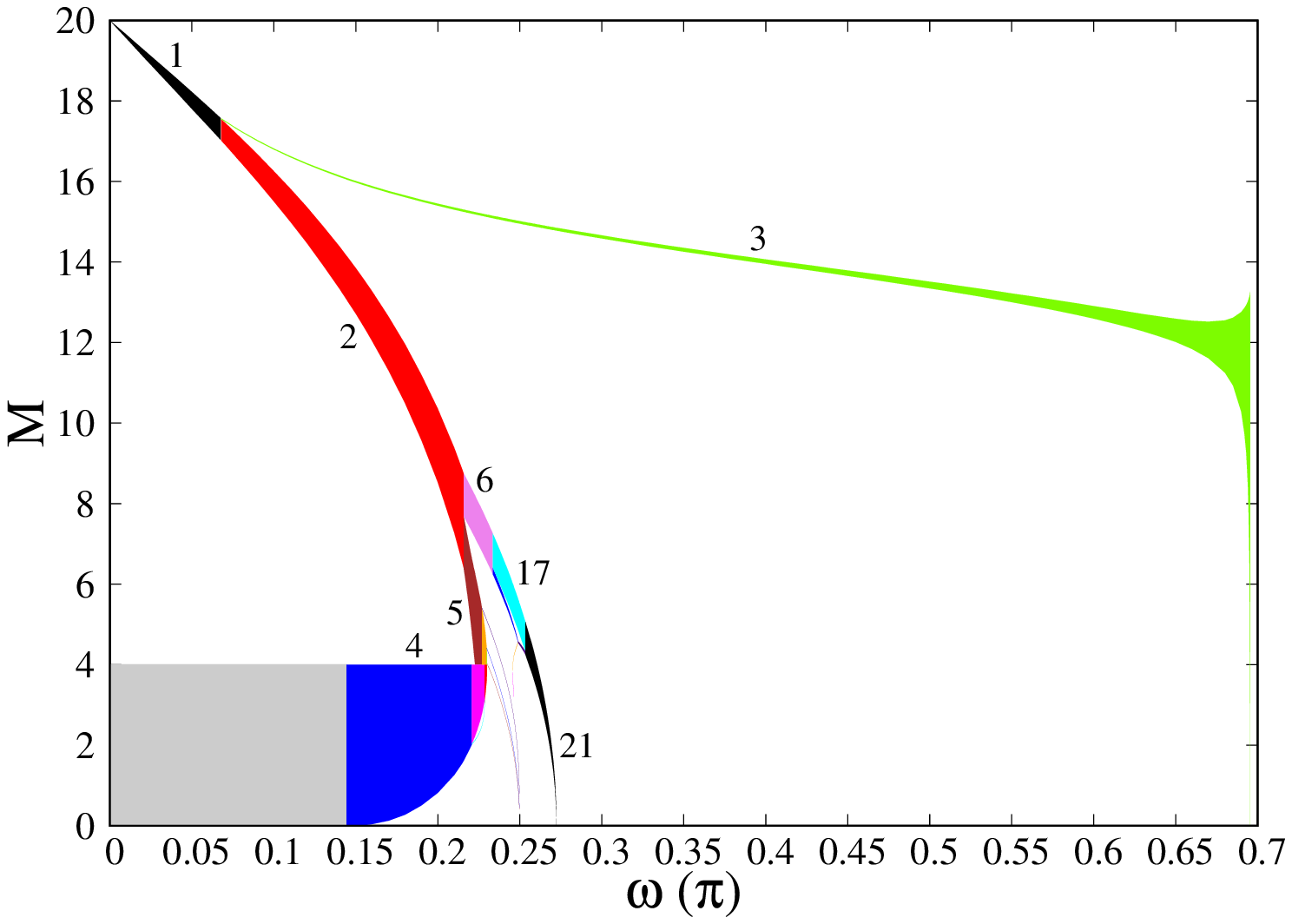}
\vspace{0pt}
\caption{Inaccessible ground-state magnetizations of Hamiltonian (\ref{eqn:model}) as a function of $\omega$. Different colors correspond to different magnetization discontinuities as they appear with increasing $\omega$, with the gray on the bottom left originating from the degeneracy of the $M=0$ and 4 ground states for small $\omega$ and zero field. Magnetization discontinuities that are clearly visible are assigned a number, in agreement with Figs \ref{fig:dodecahedrondisc} and \ref{fig:dodecahedrondisc1} (see also Fig. \ref{fig:dodecahedrondiscwidth1}).
%(~/basic/classical/dodecahedronparameter)
}
\label{fig:dodecahedrondiscwidth}
\end{figure}

\begin{figure}
\includegraphics[width=2.5in,height=3.5in,angle=270]{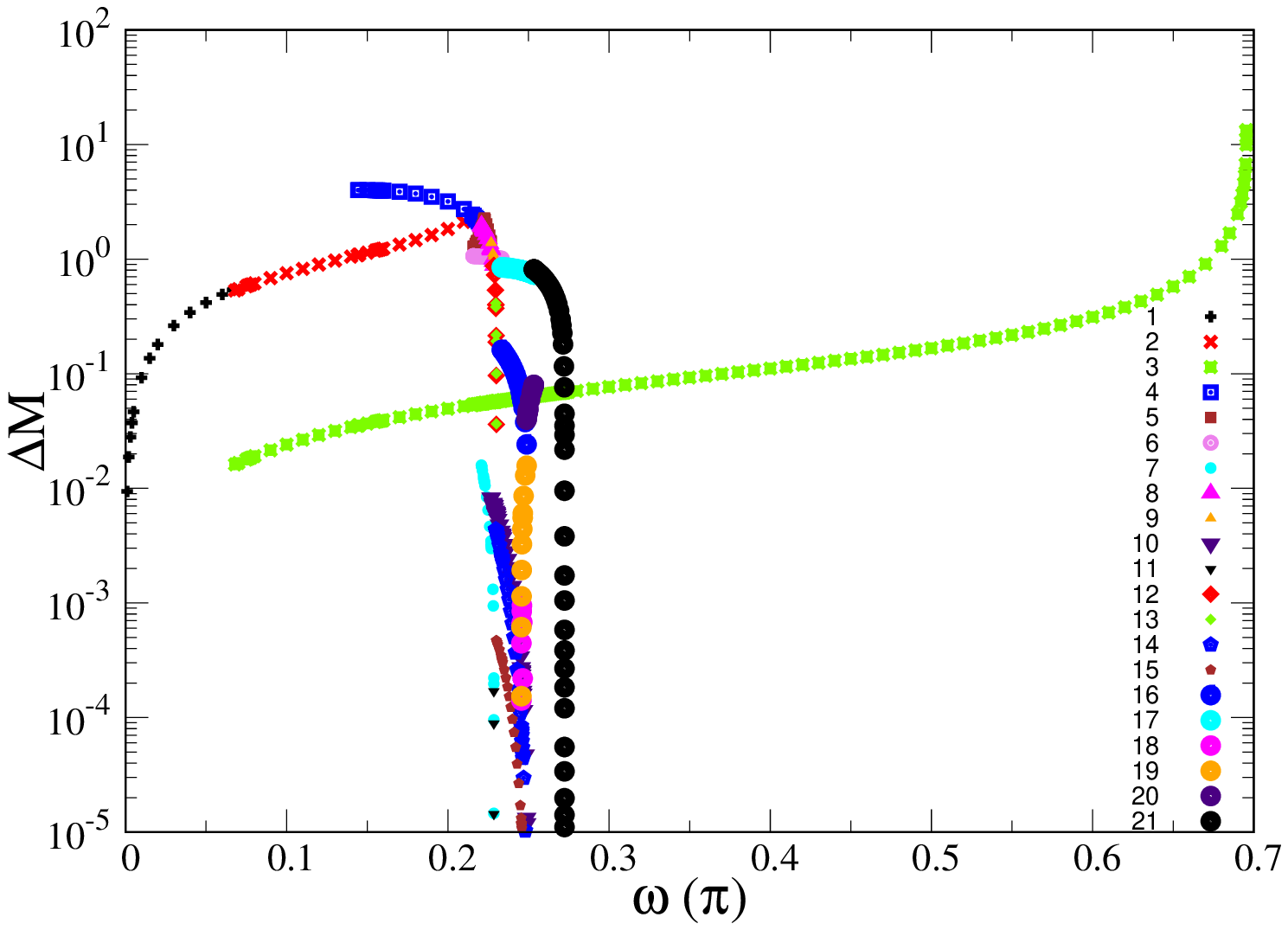}
\vspace{0pt}
\caption{Magnitude of magnetization jumps $\Delta M$ in the ground-state of Hamiltonian (\ref{eqn:model}) as a function of $\omega$. Different numbers correspond to different magnetization discontinuities according to Figs \ref{fig:dodecahedrondisc}, \ref{fig:dodecahedrondiscwidth}, \ref{fig:dodecahedrondisc1}, and \ref{fig:dodecahedrondiscwidth1}, while the colors correspond to Figs \ref{fig:dodecahedrondiscwidth} and \ref{fig:dodecahedrondiscwidth1} (see also Fig. \ref{fig:dodecahedrondiscwidth3}).
%(~/basic/classical/dodecahedronparameter)
}
\label{fig:dodecahedrondiscwidth2}
\end{figure}

\begin{figure}
\includegraphics[width=2.5in,height=3.5in,angle=270]{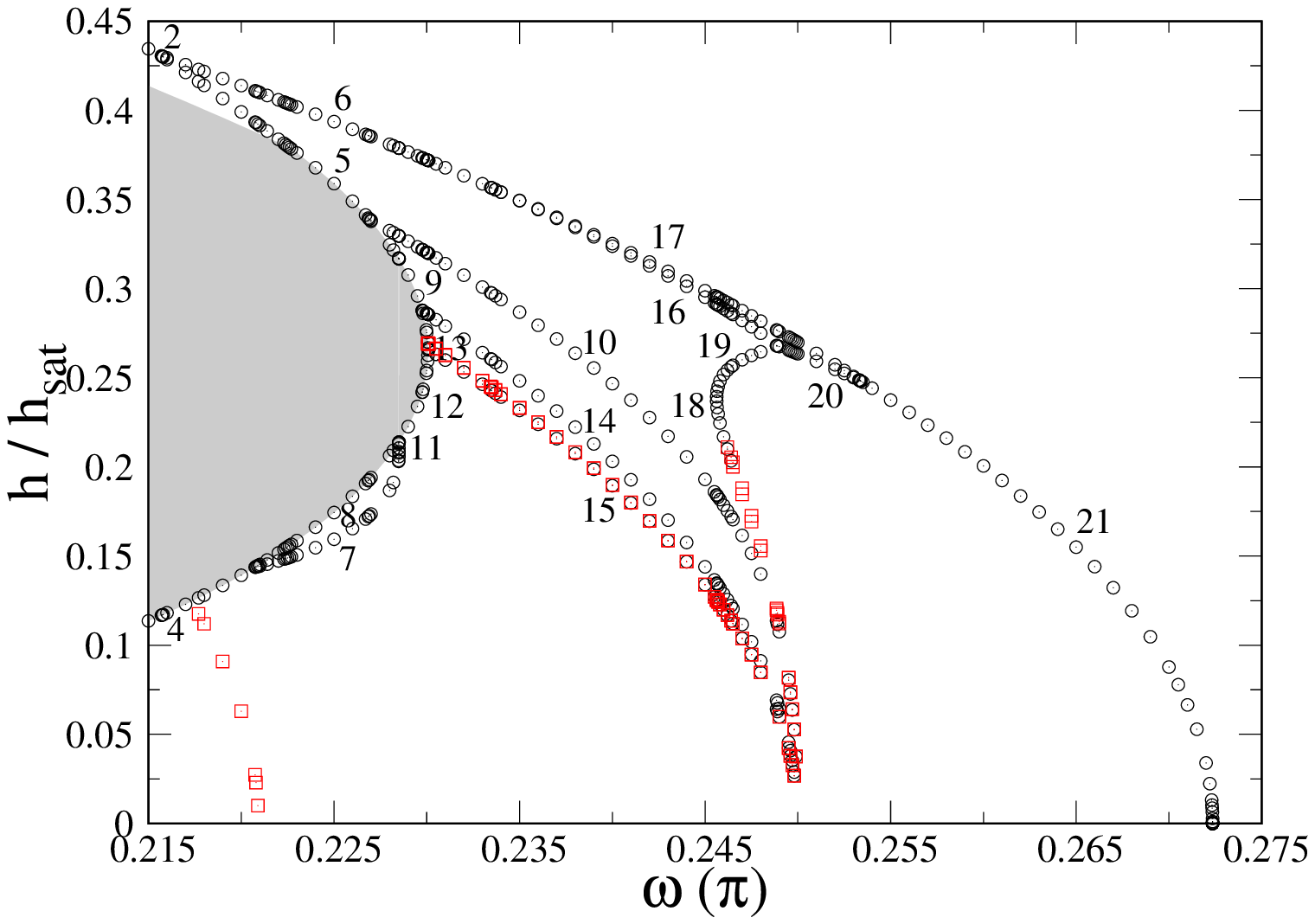}
\vspace{0pt}
\caption{Part of Fig. \ref{fig:dodecahedrondisc} in greater detail.
%(~/basic/classical/dodecahedronparameter)
}
\label{fig:dodecahedrondisc1}
\end{figure}

\begin{figure}
\includegraphics[width=2.5in,height=3.5in,angle=270]{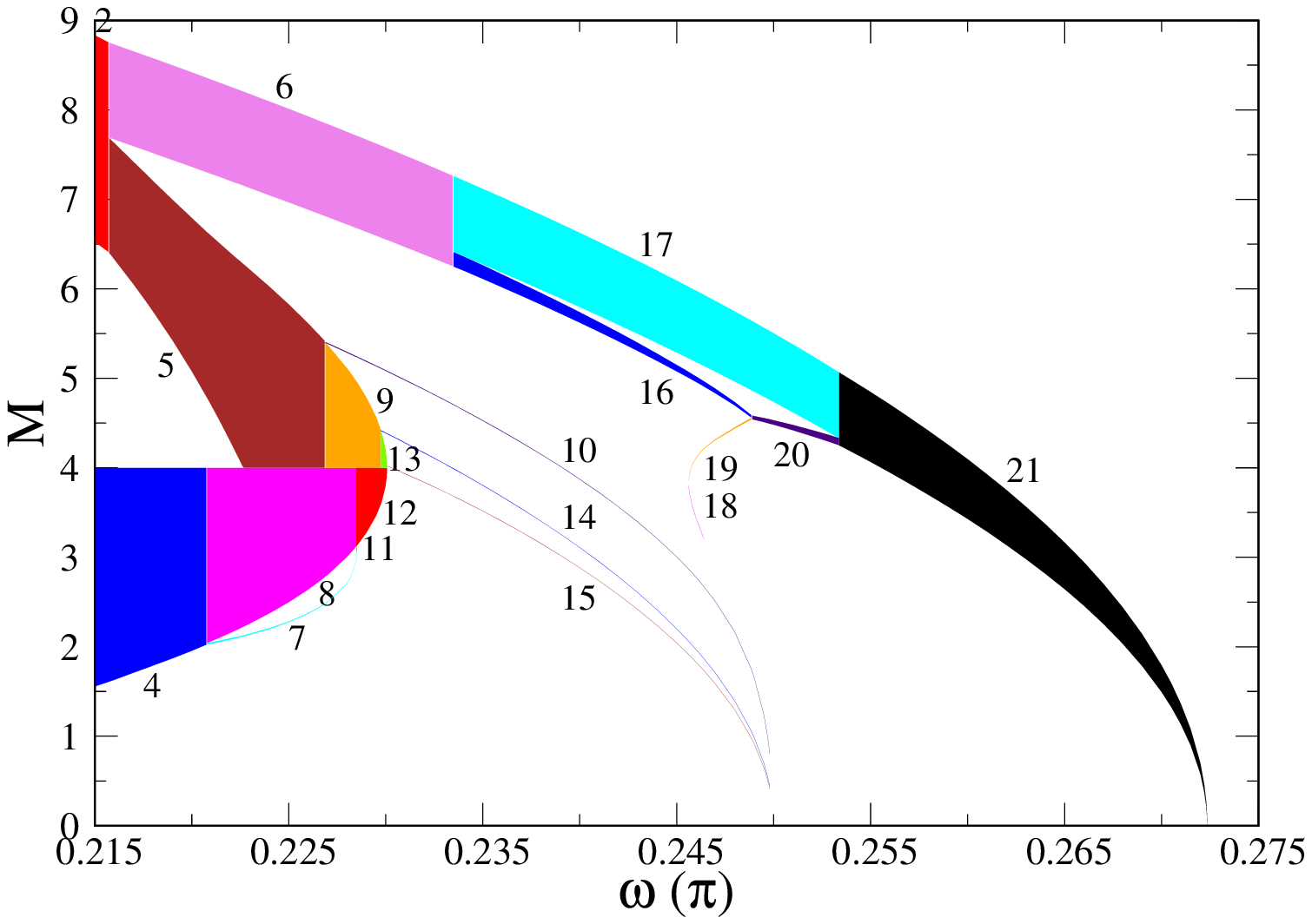}
\vspace{0pt}
\caption{Part of Fig. \ref{fig:dodecahedrondiscwidth} in greater detail.
%(~/basic/classical/dodecahedronparameter)
}
\label{fig:dodecahedrondiscwidth1}
\end{figure}

The zero-field ground-state energy develops a three-dimensional structure starting at
%$\omega=0.1443415 \pi$
$\omega=0.14434 \pi$ (Sec. \ref{sec:groundstateinzeromagneticfield}), which leads back to the $\omega=0$ Ising plateau state via a jump at a relatively small magnetic field (number 4). The plateau shrinks with $\omega$ ($\omega=0.2 \pi$ and $0.225 \pi$ in Fig. \ref{fig:dodecahedronmagnetization}) and eventually disappears at $\omega=0.23007 \pi$ (Fig. \ref{fig:dodecahedrondisc1}). A multitude of magnetization discontinuities develop for smaller magnetic fields as the plateau is about to or vanishes, with the exchange energy now also efficiently minimized in the $xy$ plane. Susceptibility discontinuities appear as well, and one of them hits the $\omega$ axis at $0.22092 \pi$ (Fig. \ref{fig:dodecahedrondisc1}). The inaccessible magnetizations are mostly confined to lower $M$ values as the isotropic Heisenberg limit $\omega=\frac{\pi}{4}$ is approached (Figs \ref{fig:dodecahedrondiscwidth} and \ref{fig:dodecahedrondiscwidth1}). The number of discontinuities for a specific $\omega$ can go up to 11, 8 of the magnetization and 3 of the susceptibility, or 7 of the magnetization and 4 of the susceptibility
%\cite{discnumber}.
\footnote{The maximum number of discontinuities for the $\omega$ values considered occur at 0.2462 and 0.2464, 8 of the magnetization and 3 of the susceptibility, and at 0.2465, 0.247, 0.2475, and 0.248, 7 of the magnetization and 4 of the susceptibility.}.

%The maximum number of discontinuities occur at 0.2462 and 0.2464, 8 of the magnetization and 3 of the susceptibility.

%The maximum number of discontinuities occur between 0.2465, 0.247, 0.2475, and 0.248, 7 of the magnetization and 4 of the susceptibility.

\begin{figure}
\includegraphics[width=2.5in,height=3.5in,angle=270]{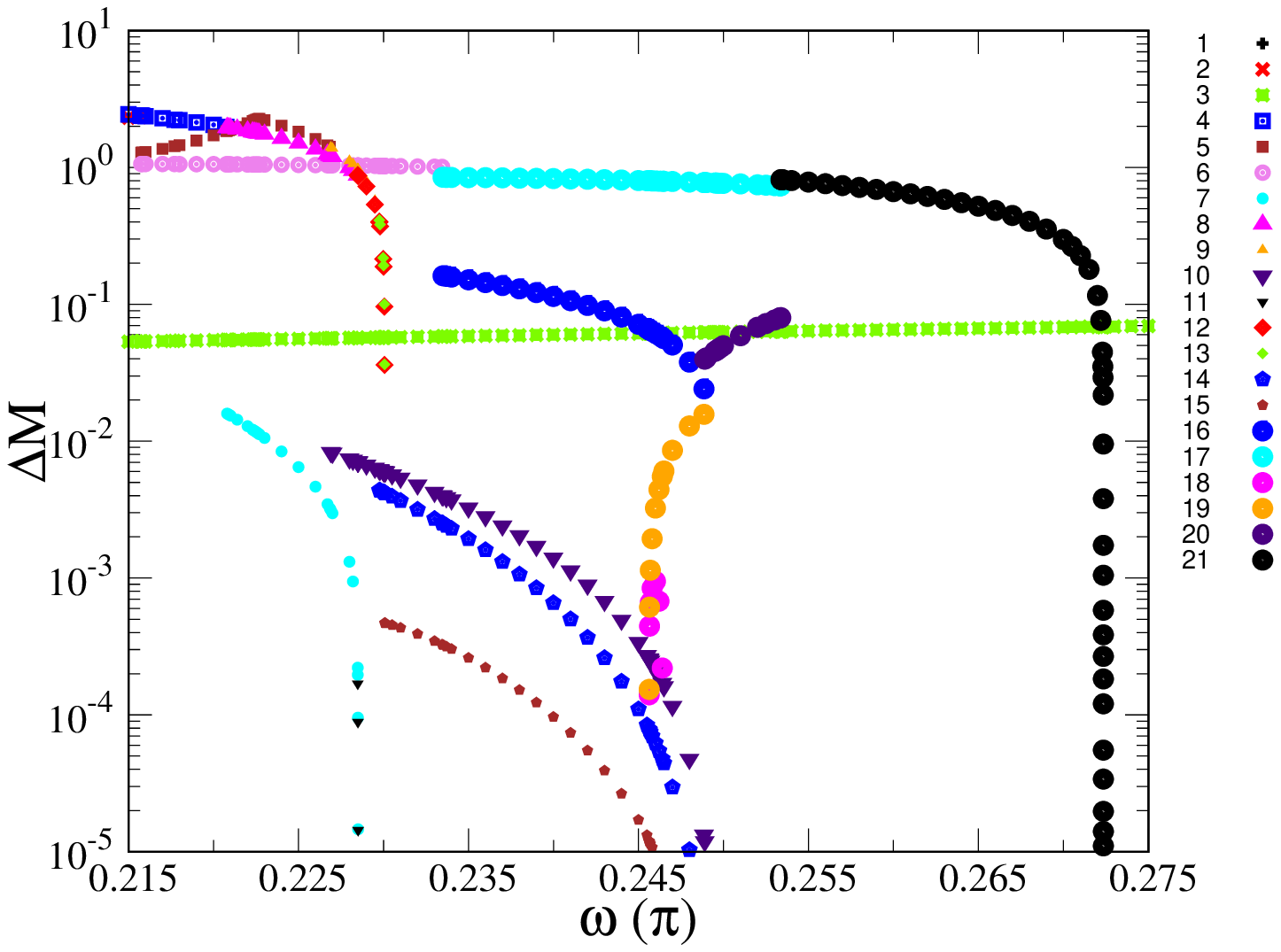}
\vspace{0pt}
\caption{Part of Fig. \ref{fig:dodecahedrondiscwidth2} in greater detail.
%(~/basic/classical/dodecahedronparameter)
}
\label{fig:dodecahedrondiscwidth3}
\end{figure}

Exactly at the AHM limit the very low-field magnetization and susceptibility discontinuities disappear. The ones that remain are the higher-field magnetization jump (number 3) and two at lower fields (numbers 17 and 20) ($\omega=0.25 \pi$ in Fig. \ref{fig:dodecahedronmagnetization}), with the three jumps occurring at $\frac{h}{h_{sat}}=0.26350$, 0.26983, and 0.73428 \cite{NPK07}.
%that can be traced to the second jump emerging from the $\omega \to 0$ high-field discontinuity and the higher field of two discontinuities that appear at $\omega=0.245(5-65) \pi$.
The two lower jumps merge at $\omega=0.25337\pi$ (number 21), and the remaining discontinuity eventually disappears at
%$\omega=0.272352 \pi$
$\omega=0.27235 \pi$ (Sec. \ref{sec:groundstateinzeromagneticfield}) at zero magnetic field (Figs \ref{fig:dodecahedrondisc1}, \ref{fig:dodecahedrondiscwidth1}, and \ref{fig:dodecahedrondiscwidth3}). The high-field magnetization discontinuity is the only one that survives at the $XY$ limit ($\omega=\frac{\pi}{2}$) at $\frac{h}{h_{sat}}=0.63679$ with $\Delta M=0.16718$. This discontinuity (shown for $\omega=0.69 \pi$ in Fig. \ref{fig:dodecahedronmagnetization}) eventually vanishes at
%$\omega=0.69550748 \pi$
$\omega=0.69551 \pi$ at zero field (Fig. \ref{fig:dodecahedrondisc}), with a large limiting value $\Delta M=13.26504$ (Figs \ref{fig:dodecahedrondiscwidth} and \ref{fig:dodecahedrondiscwidth2}) (Sec. \ref{sec:groundstateinzeromagneticfield}).

There are four lowest-energy configurations at the Heisenberg limit $\omega=\frac{\pi}{4}$ with increasing field, and they have 4, 12, 5, and 2 unique polar angles respectively \cite{NPK07}. The discontinuities separating them are 20, 17, and 3 (Figs \ref{fig:dodecahedrondisc} and \ref{fig:dodecahedrondisc1}). The high-symmetry ground state of the AHM for higher magnetic fields is of the umbrella type (App. \ref{appendix:saturationmagneticfielddodecahedron}) \cite{Landau81,Holtschneider07,NPK17-1}. It appears for infinitesimal $\omega$ just below saturation above discontinuity 1 and extends down to lower fields with increasing $\omega$ up to the ferromagnetic limit. Just below this limit it even becomes the zero-field ground state at $\omega=0.69551 \pi$ (Sec. \ref{sec:groundstateinzeromagneticfield}), as the Ising interaction is ferromagnetic for $\omega > \frac{\pi}{2}$ and supports parallel spin orientations, similarly to the magnetic field. The intermediate-field ground state of the AHM first enters for $\omega=0.06766\pi$ with discontinuity 2. This configuration becomes the zero-field ground state at $\omega=0.27235 \pi$ (Sec. \ref{sec:groundstateinzeromagneticfield}). When the field is switched on the spins start to tilt toward it, and eventually the umbrella state becomes the ground state with discontinuity 3.

\begin{figure}
\includegraphics[width=2.5in,height=3.5in,angle=270]{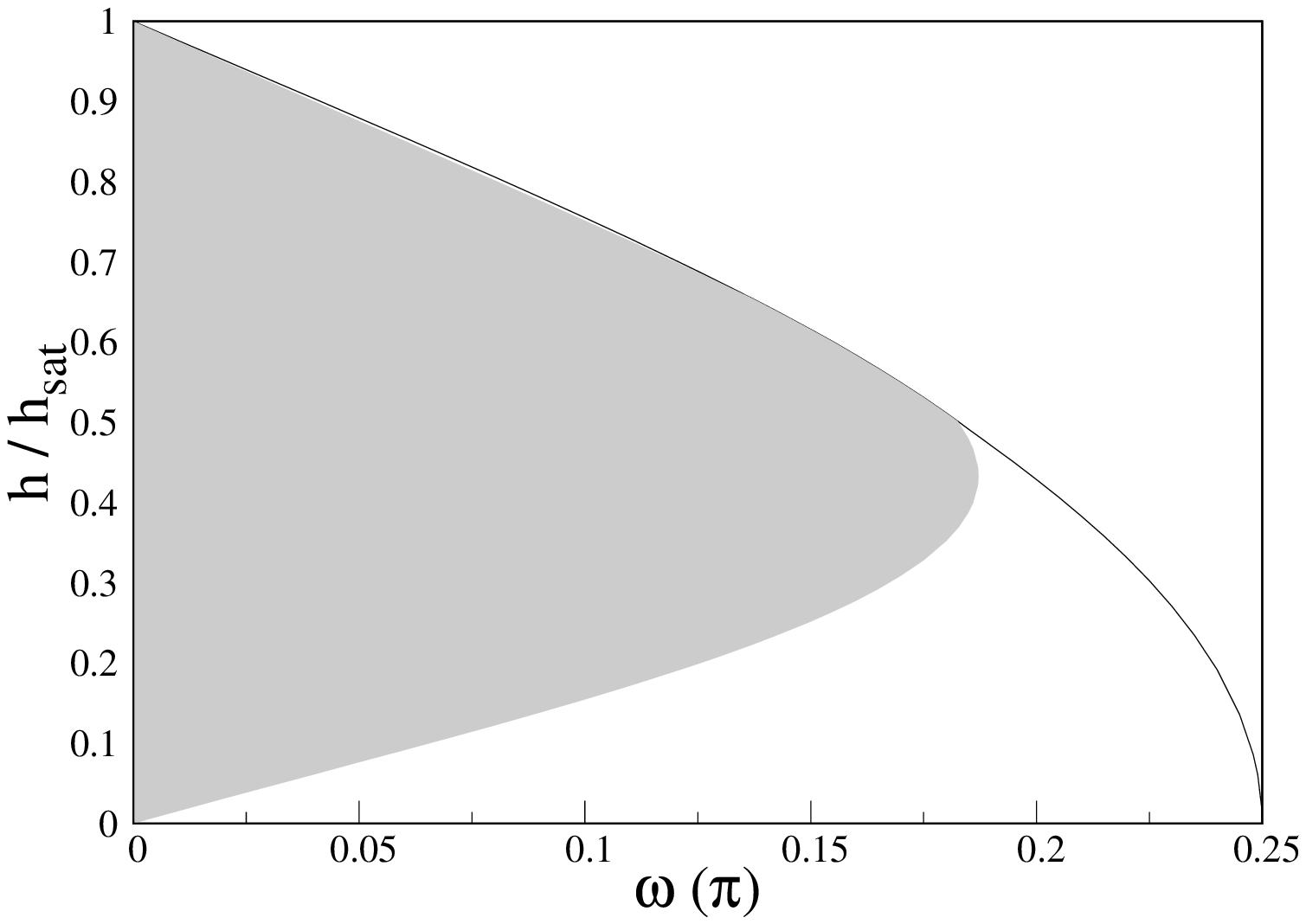}
\vspace{0pt}
\caption{Location of ground-state magnetization discontinuity (black line) and magnetization plateau (gray-shaded area) of Hamiltonian (\ref{eqn:model}) as a function of $\omega$ and the magnetic field $h$ over its saturation value $h_{sat}$ for a pentagon.
%(~/basic/classical/polygonsparameter/N=5)
}
\label{fig:pentahedrondisc}
\end{figure}

\begin{figure}
\includegraphics[width=2.5in,height=3.5in,angle=270]{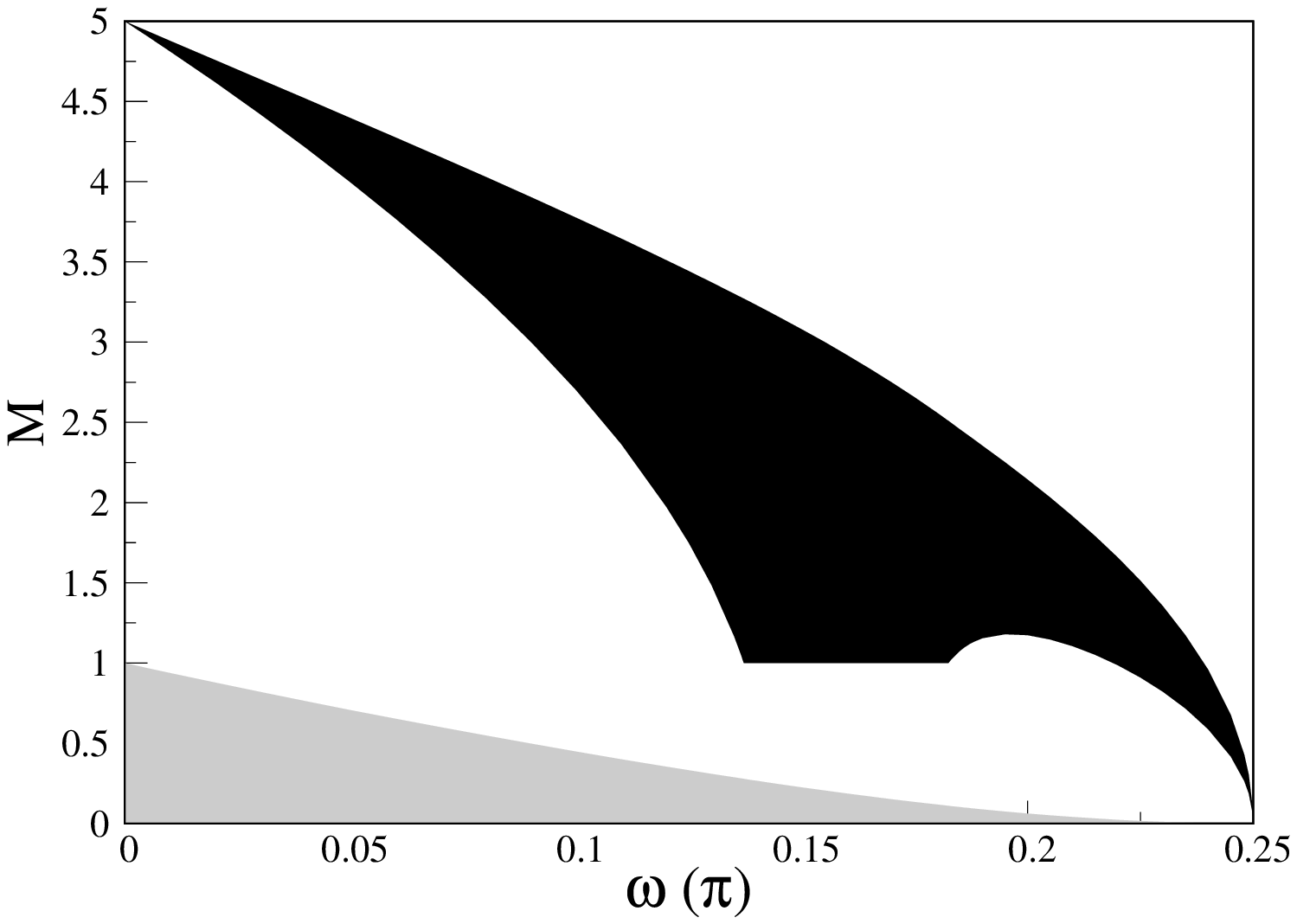}
\vspace{0pt}
\caption{Inaccessible ground-state magnetizations of Hamiltonian (\ref{eqn:model}) as a function of $\omega$ for a pentagon. The black-shaded area corresponds to the magnetization discontinuity and the gray-shaded area to the residual magnetization at zero field.
%(~/basic/classical/polygonsparameter/N=5)
}
\label{fig:pentahedrondiscwidth}
\end{figure}

The magnetic response of the dodecahedron has significant similarities with the one of its basic constituent, the pentagon, which can be viewed as an odd-numbered chain \cite{NPK15-1}. Fig. \ref{fig:pentahedrondisc} plots the location of the ground-state magnetization discontinuity and plateau as a function of $\omega$ and the magnetic field $h$ over its saturation value $h_{sat}$ for Hamiltonian (\ref{eqn:model}) on an isolated pentagon. At the Ising limit $\omega=0$ the ground state has zero-field magnetization $M=1$ forming a plateau. This exists until the magnetization becomes fully polarized with a large discontinuity $\Delta M=4$ exactly at the saturation field, just like the dodecahedron. As $\omega$ becomes finite the zero-field magnetization becomes less than 1 (Fig. \ref{fig:pentahedrondiscwidth}), with the nearest-neighbor spin correlations reducing their value now also in the $xy$ plane. The magnetization discontinuity moves away from saturation, with a wider range of magnetizations close to full polarization becoming accessible with increasing $\omega$ and the width of the plateau reduced. This has striking similarities with Figs \ref{fig:dodecahedrondisc} and \ref{fig:dodecahedrondiscwidth} and indicates that the origin of the high-field discontinuity of the dodecahedron emerging at the Ising limit lies in the magnetic response of an isolated pentagon. A further similarity is the wide magnetization range accessible following the $M=1$ plateau and leading to the discontinuity for a narrow field range. For $0.13714 < \omega < 0.18245$ the $M=1$ plateau leads directly to the discontinuity. For $\omega \geq 0.18717$ it disappears, similarly to the dodecahedron, while the discontinuity disappears exactly before the Heisenberg limit $\omega=\frac{\pi}{4}$. This shows that assembling the dodecahedron from the individual pentagons allows for the survival of the discontinuity at the Heisenberg limit and beyond.

Above the magnetization discontinuity for $\omega < \frac{\pi}{4}$ and starting at zero magnetic field for $\omega \geq \frac{\pi}{4}$ the ground state is of the umbrella type, again resembling the dodecahedron. This includes the $XY$ limit $\omega=\frac{\pi}{2}$. However, the dodecahedron admits an umbrella-type zero-field ground state only very close to the ferromagnetic limit, demonstrating the importance of its connectivity in its magnetization response. At $\omega = \pi - tan^{-1}(\sqrt{5}-1)$ the ground state becomes ferromagnetic along the Ising axis even at zero field, very close to the corresponding value of the dodecahedron. The saturation magnetic field $h_{sat} = 2cos\omega + \frac{\sqrt{5}+1}{2} sin\omega$ (App. \ref{appendix:saturationmagneticfieldpentagon}).

\begin{figure}
\includegraphics[width=2.5in,height=3.5in,angle=270]{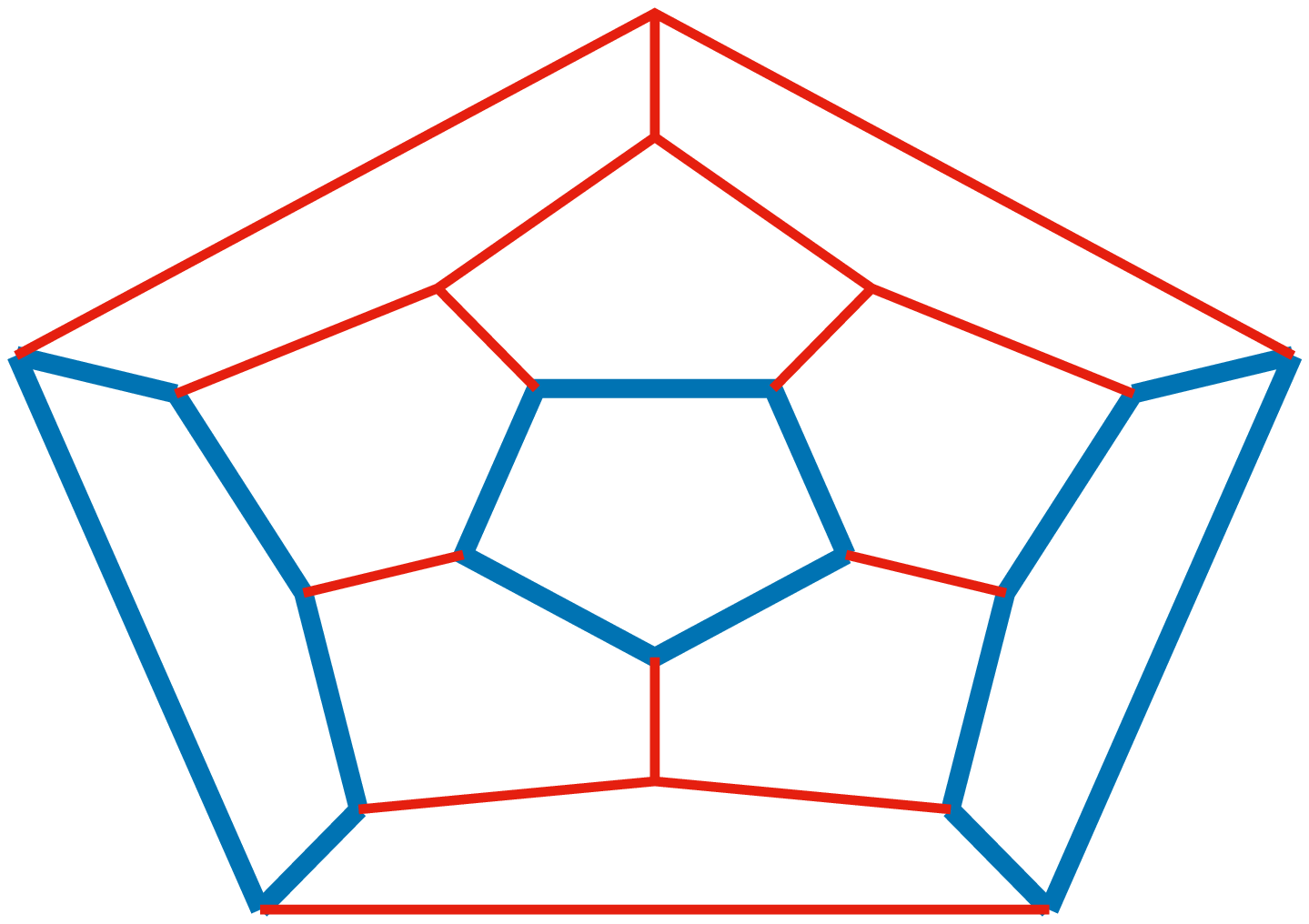}
\vspace{0pt}
\caption{The thick (blue) lines show three isolated pentagons in the planar projection of the dodecahedron. The thin (red) lines show the rest of the bonds. The thick bonds have strength $cos\phi$ and the thin bonds $sin\phi$.
%(~/basic/classical/fullerenes)
}
\label{fig:pentagonsisolated}
\end{figure}

The transition from the isolated pentagon to the dodecahedron limit can be effected by selecting three nonneighboring pentagons (Fig. \ref{fig:pentagonsisolated}), and then allowing the rest of the dodecahedron couplings to increase from zero until all the couplings are equal. The bonds within the three pentagons are taken to have strength $cos\phi$ while the rest $sin\phi$, with $0 \leq \phi \leq \frac{\pi}{4}$. The evolution of the magnetization and susceptibility discontinuities as a function of the parameter $\phi$ that controls the couplings is shown in Fig. \ref{fig:pentagonsisolateddisc} for the AHM limit $\omega=\frac{\pi}{4}$. The umbrella-type ground state of the isolated pentagon limit $\phi=0$ directly evolves to the high-field phase of the dodecahedron limit $\phi=\frac{\pi}{4}$, demonstrating again that the origin of the high-field magnetization response of the dodecahedron reflects the magnetic behavior of an isolated pentagon.

It is noted that Hamiltonian (\ref{eqn:model}) on an isolated triangle is not associated with a discontinuity but only with a magnetization plateau. It has been shown that the mechanism of the magnetization discontinuity of the AHM on the icosahedron is related to its structure being equivalent to a closed strip of a triangular lattice with two additional spins attached. The discontinuity evolves continuously from the one effected by these two spins when they are uncoupled to the cluster \cite{NPK15}.

\begin{figure}
\includegraphics[width=3.5in,height=2.5in]{dodecahedronpentagons}
\vspace{0pt}
\caption{Location of ground-state magnetization (black circles) and susceptibility (red squares) discontinuities of the AHM for $0 \leq \phi \leq \frac{\pi}{4}$ (Fig. \ref{fig:pentagonsisolated}) as a function of $\phi$ and the magnetic field $h$ over its saturation value $h_{sat}$ for the dodecahedron.
%(~/basic/classical/dodecahedronpentagons)
}
\label{fig:pentagonsisolateddisc}
\end{figure}

\section{Conclusions}
\label{sec:conclusions}

The classical antiferromagnetic Heisenberg model on the dodecahedron has three magnetization discontinuities in an external field \cite{NPK07}. In this paper it was shown how these discontinuities originate from the magnetic response at the Ising limit, which consists of an extended magnetization plateau and a jump to saturation, due to the magnetic response of an isolated pentagon and the frustrated connectivity of the dodecahedron.
%This is similar to what occurs in the quantum-mechanical case \cite{NPK23-1}.
The highest-field discontinuity persists way after the antiferromagnetic $XY$ limit, and vanishes just before the ferromagnetic Ising interaction aligns all the spins. The AHM on other icosahedral fullerenes has similar magnetic response with the one of the dodecahedron \cite{NPK07}, further demonstrating the importance of an isolated pentagon for the existence of magnetization discontinuities in molecules that include this polygon.

In the quantum-mechanical case the discontinuity five spin flips away from saturation also originates from the Ising-limit discontinuity and appears once the interaction in the $xy$ plane becomes finite. It does not vanish even at the Heisenberg limit for individual spins $s \leq 1$. However the discontinuity disappears closer to the Ising limit with increasing $s$ (Fig. 7 and Table 5 in Ref. \cite{NPK23-1}). This clearly distinguishes the quantum case of finite $s$ with the classical case where $s \to \infty$ and the discontinuity reappears. Ref. \cite{NPK01} calculates the quantum ground state by perturbing with quantum fluctuations away from coherent states derived from the classical ground state.

The findings in this paper demonstrate that the calculation of the magnetic response of the Ising model can be of particular use for frustrated molecules. The existence of Ising discontinuities is an indication that they may survive at the Heisenberg limit, either at the classical or the quantum level. This is even more so when the Ising jumps are of significant strength, as is the case for the dodecahedron. At the same time, the calculation of the Ising magnetization curve can be done relatively quickly. Such an approach has already been useful for the AHM on the dodecahedron as well as the truncated icosahedron, a molecule also of icosahedral symmetry \cite{NPK23-1}. It is of even bigger importance when the discontinuities can not be directly explained by the connectivity of the molecule \cite{NPK17}, as was done for the icosahedron \cite{NPK15}.

\begin{appendix}

\section{Saturation Magnetic Field for the Dodecahedron}
\label{appendix:saturationmagneticfielddodecahedron}

When $0 \leq \omega \leq \pi - tan^{-1}\frac{3}{\sqrt{5}}$ in the lowest-energy configuration just below saturation the spins assume two distinct polar angles $\theta_1$ and $\theta_2$ each corresponding to ten spins, while the azimuthal angles acquire ten different values that differ by $\frac{\pi}{5}$. The nearest-neighbor correlations assume three distinct values and the energy functional is

\begin{eqnarray}
\frac{E}{10} & = & sin\omega sin^2\theta_1 cos\frac{4\pi}{5} + cos\omega cos^2\theta_1 + \nonumber \\ & & sin\omega sin^2\theta_2 cos\frac{3\pi}{5} + cos\omega cos^2\theta_2 - \nonumber \\ & & sin\omega sin\theta_1 sin\theta_2 + cos\omega cos\theta_1 cos\theta_2 - \nonumber \\ & & h ( cos\theta_1 + cos\theta_2 )
\end{eqnarray}

This eventually becomes

\begin{eqnarray}
\frac{E}{10} - 2 cos\omega & = & - ( \frac{\sqrt{5}+1}{4} sin\omega + cos\omega ) sin^2\theta_1 - \nonumber \\ & & ( \frac{\sqrt{5}-1}{4} sin\omega + cos\omega ) sin^2\theta_2 - \nonumber \\ & & sin\omega sin\theta_1 sin\theta_2 + cos\omega cos\theta_1 cos\theta_2 - \nonumber \\ & & h ( cos\theta_1 + cos\theta_2 )
\end{eqnarray}

Close to saturation $\theta_1 \to 0$ and $\theta_2 \to 0$, and a small angle expansion gives

\begin{eqnarray}
\frac{E}{5} - 6 cos\omega + 4h & \approx & - ( \frac{\sqrt{5}+1}{2} sin\omega + 3 cos\omega - h ) \theta^2_1 - \nonumber \\ & & ( \frac{\sqrt{5}-1}{2} sin\omega + 3 cos\omega - h ) \theta^2_2 - \nonumber \\ & & 2 sin\omega \theta_1 \theta_2
\end{eqnarray}

The derivatives with respect to the two unique polar angles are
\begin{eqnarray}
\frac{\partial (\frac{E}{5} - 6 cos\omega + 4h)}{\partial \theta_1} & \approx & - [ (\sqrt{5}+1) sin\omega + 6 cos\omega - 2 h ] \theta_1 - \nonumber \\ & & 2 sin\omega \theta_2 \nonumber \\
\frac{\partial (\frac{E}{5} - 6 cos\omega + 4h)}{\partial \theta_2} & \approx & - [ (\sqrt{5}-1) sin\omega + 6 cos\omega - 2 h ] \theta_2 - \nonumber \\ & & 2 sin\omega \theta_1
\end{eqnarray}

To find the minimum the derivatives are set equal to zero, leading to the equation $h^2-(\sqrt{5} sin\omega +6 cos\omega)h+9 cos^2\omega+3\sqrt{5} sin\omega cos\omega=0$. This gives for the saturation field $h_{sat}=3cos\omega + \sqrt{5} sin\omega$.

\section{Saturation Magnetic Field for a Pentagon}
\label{appendix:saturationmagneticfieldpentagon}

When $0 \leq \omega \leq \pi - tan^{-1}(\sqrt{5}-1)$ in the lowest-energy configuration just below saturation all spins share a polar angle $\theta$, while nearest-neighbor spins form an angle of $\frac{4\pi}{5}$ in the azimuthal plane. The nearest-neighbor correlations assume a single value and the energy functional is

\begin{eqnarray}
& & \frac{E}{5} = sin\omega sin^2\theta cos\frac{4\pi}{5} + cos\omega cos^2\theta - h cos\theta
\end{eqnarray}

This eventually becomes

\begin{eqnarray}
& & \frac{E}{5} + \frac{\sqrt{5}+1}{4} sin\omega = (\frac{\sqrt{5}+1}{4} sin\omega+cos\omega) cos^2\theta - \nonumber \\ & & h cos\theta
\end{eqnarray}

The derivative with respect to $\theta$ is

\begin{eqnarray}
\frac{\partial(\frac{E}{5} + \frac{\sqrt{5}+1}{4} sin\omega)}{\partial \theta} & = & sin\theta [-(\frac{\sqrt{5}+1}{2} sin\omega+2cos\omega) cos\theta + \nonumber \\ & & h ]
\end{eqnarray}

To find the minimum the derivative is set equal to zero. This gives for the saturation field $h_{sat} = 2cos\omega + \frac{\sqrt{5}+1}{2} sin\omega$.

\end{appendix}

\bibliography{dodecahedronparameter}

%merlin.mbs apsrev4-1.bst 2010-07-25 4.21a (PWD, AO, DPC) hacked
%Control: key (0)
%Control: author (8) initials jnrlst
%Control: editor formatted (1) identically to author
%Control: production of article title (-1) disabled
%Control: page (0) single
%Control: year (1) truncated
%Control: production of eprint (0) enabled
\begin{thebibliography}{46}%
\makeatletter
\providecommand \@ifxundefined [1]{%
 \@ifx{#1\undefined}
}%
\providecommand \@ifnum [1]{%
 \ifnum #1\expandafter \@firstoftwo
 \else \expandafter \@secondoftwo
 \fi
}%
\providecommand \@ifx [1]{%
 \ifx #1\expandafter \@firstoftwo
 \else \expandafter \@secondoftwo
 \fi
}%
\providecommand \natexlab [1]{#1}%
\providecommand \enquote  [1]{``#1''}%
\providecommand \bibnamefont  [1]{#1}%
\providecommand \bibfnamefont [1]{#1}%
\providecommand \citenamefont [1]{#1}%
\providecommand \href@noop [0]{\@secondoftwo}%
\providecommand \href [0]{\begingroup \@sanitize@url \@href}%
\providecommand \@href[1]{\@@startlink{#1}\@@href}%
\providecommand \@@href[1]{\endgroup#1\@@endlink}%
\providecommand \@sanitize@url [0]{\catcode `\\12\catcode `\$12\catcode
  `\&12\catcode `\#12\catcode `\^12\catcode `\_12\catcode `\%12\relax}%
\providecommand \@@startlink[1]{}%
\providecommand \@@endlink[0]{}%
\providecommand \url  [0]{\begingroup\@sanitize@url \@url }%
\providecommand \@url [1]{\endgroup\@href {#1}{\urlprefix }}%
\providecommand \urlprefix  [0]{URL }%
\providecommand \Eprint [0]{\href }%
\providecommand \doibase [0]{http://dx.doi.org/}%
\providecommand \selectlanguage [0]{\@gobble}%
\providecommand \bibinfo  [0]{\@secondoftwo}%
\providecommand \bibfield  [0]{\@secondoftwo}%
\providecommand \translation [1]{[#1]}%
\providecommand \BibitemOpen [0]{}%
\providecommand \bibitemStop [0]{}%
\providecommand \bibitemNoStop [0]{.\EOS\space}%
\providecommand \EOS [0]{\spacefactor3000\relax}%
\providecommand \BibitemShut  [1]{\csname bibitem#1\endcsname}%
\let\auto@bib@innerbib\@empty
%</preamble>
\bibitem [{\citenamefont {Plato}()}]{Plato}%
  \BibitemOpen
  \bibfield  {author} {\bibinfo {author} {\bibnamefont {Plato}},\ }\href@noop
  {} {\bibinfo  {journal} {{\it Timaeus}}\ }\BibitemShut {NoStop}%
\bibitem [{\citenamefont {Goldberg}(1937)}]{Goldberg37}%
  \BibitemOpen
\bibfield  {journal} {  }\bibfield  {author} {\bibinfo {author} {\bibfnamefont
  {M.}~\bibnamefont {Goldberg}},\ }\href@noop {} {\bibfield  {journal}
  {\bibinfo  {journal} {Tohoku Math. J.}\ }\textbf {\bibinfo {volume} {43}},\
  \bibinfo {pages} {104} (\bibinfo {year} {1937})}\BibitemShut {NoStop}%
\bibitem [{\citenamefont {Fowler}\ and\ \citenamefont
  {Manolopoulos}(1995)}]{Fowler95}%
  \BibitemOpen
  \bibfield  {author} {\bibinfo {author} {\bibfnamefont {P.~W.}\ \bibnamefont
  {Fowler}}\ and\ \bibinfo {author} {\bibfnamefont {D.~E.}\ \bibnamefont
  {Manolopoulos}},\ }\href@noop {} {\emph {\bibinfo {title} {An Atlas of
  Fullerenes}}}\ (\bibinfo  {publisher} {Oxford University Press},\ \bibinfo
  {address} {Oxford},\ \bibinfo {year} {1995})\BibitemShut {NoStop}%
\bibitem [{\citenamefont {Prinzbach}\ \emph {et~al.}(2000)\citenamefont
  {Prinzbach}, \citenamefont {Weiler}, \citenamefont {Landenberger} \emph
  {et~al.}}]{Prinzbach00}%
  \BibitemOpen
  \bibfield  {author} {\bibinfo {author} {\bibfnamefont {H.}~\bibnamefont
  {Prinzbach}}, \bibinfo {author} {\bibfnamefont {A.}~\bibnamefont {Weiler}},
  \bibinfo {author} {\bibfnamefont {P.}~\bibnamefont {Landenberger}},  \emph
  {et~al.},\ }\href {\doibase 10.1038/35024037} {\bibfield  {journal} {\bibinfo
   {journal} {Nature}\ }\textbf {\bibinfo {volume} {407}},\ \bibinfo {pages}
  {60} (\bibinfo {year} {2000})}\BibitemShut {NoStop}%
\bibitem [{\citenamefont {Wang}\ \emph {et~al.}(2001)\citenamefont {Wang},
  \citenamefont {Ke}, \citenamefont {Zhu} \emph {et~al.}}]{Wang01}%
  \BibitemOpen
  \bibfield  {author} {\bibinfo {author} {\bibfnamefont {Z.}~\bibnamefont
  {Wang}}, \bibinfo {author} {\bibfnamefont {X.}~\bibnamefont {Ke}}, \bibinfo
  {author} {\bibfnamefont {Z.}~\bibnamefont {Zhu}},  \emph {et~al.},\ }\href
  {\doibase 10.1016/S0375-9601(00)00847-1} {\bibfield  {journal} {\bibinfo
  {journal} {Phys. Lett. A}\ }\textbf {\bibinfo {volume} {280}},\ \bibinfo
  {pages} {351} (\bibinfo {year} {2001})}\BibitemShut {NoStop}%
\bibitem [{\citenamefont {Iqbal}\ \emph {et~al.}(2003)\citenamefont {Iqbal},
  \citenamefont {Zhang}, \citenamefont {Grebel} \emph {et~al.}}]{Iqbal03}%
  \BibitemOpen
  \bibfield  {author} {\bibinfo {author} {\bibfnamefont {Z.}~\bibnamefont
  {Iqbal}}, \bibinfo {author} {\bibfnamefont {Y.}~\bibnamefont {Zhang}},
  \bibinfo {author} {\bibfnamefont {H.}~\bibnamefont {Grebel}},  \emph
  {et~al.},\ }\href {\doibase 10.1140/epjb/e2003-00060-4} {\bibfield  {journal}
  {\bibinfo  {journal} {Eur. Phys. J. B}\ }\textbf {\bibinfo {volume} {31}},\
  \bibinfo {pages} {509} (\bibinfo {year} {2003})}\BibitemShut {NoStop}%
\bibitem [{\citenamefont {Qin}\ \emph {et~al.}(2017)\citenamefont {Qin},
  \citenamefont {Zhou}, \citenamefont {Yu} \emph {et~al.}}]{Qin17}%
  \BibitemOpen
  \bibfield  {author} {\bibinfo {author} {\bibfnamefont {L.}~\bibnamefont
  {Qin}}, \bibinfo {author} {\bibfnamefont {G.-J.}\ \bibnamefont {Zhou}},
  \bibinfo {author} {\bibfnamefont {Y.-Z.}\ \bibnamefont {Yu}},  \emph
  {et~al.},\ }\href {\doibase https://doi.org/10.1021/jacs.7b09996} {\bibfield
  {journal} {\bibinfo  {journal} {J. Am. Chem. Soc.}\ }\textbf {\bibinfo
  {volume} {139}},\ \bibinfo {pages} {16405} (\bibinfo {year}
  {2017})}\BibitemShut {NoStop}%
\bibitem [{\citenamefont {Altmann}\ and\ \citenamefont
  {Herzig}(1994)}]{Altmann94}%
  \BibitemOpen
  \bibfield  {author} {\bibinfo {author} {\bibfnamefont {S.~L.}\ \bibnamefont
  {Altmann}}\ and\ \bibinfo {author} {\bibfnamefont {P.}~\bibnamefont
  {Herzig}},\ }\href@noop {} {\emph {\bibinfo {title} {Point-Group Theory
  Tables}}}\ (\bibinfo  {publisher} {Oxford University Press},\ \bibinfo
  {address} {London},\ \bibinfo {year} {1994})\BibitemShut {NoStop}%
\bibitem [{\citenamefont {Auerbach}(1998)}]{Auerbach98}%
  \BibitemOpen
  \bibfield  {author} {\bibinfo {author} {\bibfnamefont {A.}~\bibnamefont
  {Auerbach}},\ }\href {\doibase 10.1007/978-1-4612-0869-3} {\emph {\bibinfo
  {title} {Interacting Electrons and Quantum Magnetism}}}\ (\bibinfo
  {publisher} {Springer Verlag},\ \bibinfo {address} {New York},\ \bibinfo
  {year} {1998})\BibitemShut {NoStop}%
\bibitem [{\citenamefont {Fazekas}(1999)}]{Fazekas99}%
  \BibitemOpen
  \bibfield  {author} {\bibinfo {author} {\bibfnamefont {P.}~\bibnamefont
  {Fazekas}},\ }\href {\doibase 10.1142/2945} {\emph {\bibinfo {title} {Lecture
  Notes on Electron Correlation and Magnetism}}}\ (\bibinfo  {publisher} {World
  Scientific},\ \bibinfo {address} {Singapore},\ \bibinfo {year}
  {1999})\BibitemShut {NoStop}%
\bibitem [{\citenamefont {Lhuillier}\ and\ \citenamefont
  {Misguich}(2001)}]{Lhuillier01}%
  \BibitemOpen
  \bibfield  {author} {\bibinfo {author} {\bibfnamefont {C.}~\bibnamefont
  {Lhuillier}}\ and\ \bibinfo {author} {\bibfnamefont {G.}~\bibnamefont
  {Misguich}},\ }\href@noop {} {\emph {\bibinfo {title} {{\normalfont in} High
  Magnetic Fields Applications in Condensed Matter Physics and Spectroscopy,
  {\normalfont edited by C. Berthier, L. P. Levy, and G. Martinez}}}}\
  (\bibinfo  {publisher} {Springer},\ \bibinfo {address} {New York},\ \bibinfo
  {year} {2001})\BibitemShut {NoStop}%
\bibitem [{\citenamefont {Misguich}\ and\ \citenamefont
  {Lhuillier}(2003)}]{Misguich03}%
  \BibitemOpen
  \bibfield  {author} {\bibinfo {author} {\bibfnamefont {G.}~\bibnamefont
  {Misguich}}\ and\ \bibinfo {author} {\bibfnamefont {C.}~\bibnamefont
  {Lhuillier}},\ }\href@noop {} {\emph {\bibinfo {title} {{\normalfont in}
  Frustrated Spin Systems, {\normalfont edited by H. T. Diep}}}}\ (\bibinfo
  {publisher} {World Scientific},\ \bibinfo {address} {Singapore},\ \bibinfo
  {year} {2003})\BibitemShut {NoStop}%
\bibitem [{\citenamefont {Ramirez}(2005)}]{Ramirez05}%
  \BibitemOpen
  \bibfield  {author} {\bibinfo {author} {\bibfnamefont {A.~P.}\ \bibnamefont
  {Ramirez}},\ }\href {\doibase 10.1557/mrs2005.122} {\bibfield  {journal}
  {\bibinfo  {journal} {MRS Bull.}\ }\textbf {\bibinfo {volume} {30}},\
  \bibinfo {pages} {447} (\bibinfo {year} {2005})}\BibitemShut {NoStop}%
\bibitem [{\citenamefont {Schnack}(2010)}]{Schnack10}%
  \BibitemOpen
  \bibfield  {author} {\bibinfo {author} {\bibfnamefont {J.}~\bibnamefont
  {Schnack}},\ }\href {\doibase 10.1039/B925358K} {\bibfield  {journal}
  {\bibinfo  {journal} {Dalton Trans.}\ }\textbf {\bibinfo {volume} {39}},\
  \bibinfo {pages} {4677} (\bibinfo {year} {2010})}\BibitemShut {NoStop}%
\bibitem [{\citenamefont {Schmidt}\ and\ \citenamefont
  {Luban}(2003)}]{Schmidt03}%
  \BibitemOpen
  \bibfield  {author} {\bibinfo {author} {\bibfnamefont {H.-J.}\ \bibnamefont
  {Schmidt}}\ and\ \bibinfo {author} {\bibfnamefont {M.}~\bibnamefont
  {Luban}},\ }\href {\doibase 10.1088/0305-4470/36/23/306} {\bibfield
  {journal} {\bibinfo  {journal} {J. Phys. A: Math. and Gen.}\ }\textbf
  {\bibinfo {volume} {36}},\ \bibinfo {pages} {6351} (\bibinfo {year}
  {2003})}\BibitemShut {NoStop}%
\bibitem [{\citenamefont {Coffey}\ and\ \citenamefont
  {Trugman}(1992)}]{Coffey92}%
  \BibitemOpen
  \bibfield  {author} {\bibinfo {author} {\bibfnamefont {D.}~\bibnamefont
  {Coffey}}\ and\ \bibinfo {author} {\bibfnamefont {S.~A.}\ \bibnamefont
  {Trugman}},\ }\href {\doibase 10.1103/PhysRevLett.69.176} {\bibfield
  {journal} {\bibinfo  {journal} {Phys. Rev. Lett.}\ }\textbf {\bibinfo
  {volume} {69}},\ \bibinfo {pages} {176} (\bibinfo {year} {1992})}\BibitemShut
  {NoStop}%
\bibitem [{\citenamefont {Konstantinidis}(2007)}]{NPK07}%
  \BibitemOpen
  \bibfield  {author} {\bibinfo {author} {\bibfnamefont {N.~P.}\ \bibnamefont
  {Konstantinidis}},\ }\href {\doibase 10.1103/PhysRevB.76.104434} {\bibfield
  {journal} {\bibinfo  {journal} {Phys. Rev. B}\ }\textbf {\bibinfo {volume}
  {76}},\ \bibinfo {pages} {104434} (\bibinfo {year} {2007})}\BibitemShut
  {NoStop}%
\bibitem [{\citenamefont {Konstantinidis}(2005)}]{NPK05}%
  \BibitemOpen
  \bibfield  {author} {\bibinfo {author} {\bibfnamefont {N.~P.}\ \bibnamefont
  {Konstantinidis}},\ }\href {\doibase 10.1103/PhysRevB.72.064453} {\bibfield
  {journal} {\bibinfo  {journal} {Phys. Rev. B}\ }\textbf {\bibinfo {volume}
  {72}},\ \bibinfo {pages} {064453} (\bibinfo {year} {2005})}\BibitemShut
  {NoStop}%
\bibitem [{\citenamefont {Konstantinidis}(2023{\natexlab{a}})}]{NPK23-1}%
  \BibitemOpen
  \bibfield  {author} {\bibinfo {author} {\bibfnamefont {N.~P.}\ \bibnamefont
  {Konstantinidis}},\ }\href {\doibase 10.21468/SciPostPhys.15.1.037}
  {\bibfield  {journal} {\bibinfo  {journal} {SciPost Phys.}\ }\textbf
  {\bibinfo {volume} {15}},\ \bibinfo {pages} {037} (\bibinfo {year}
  {2023}{\natexlab{a}})}\BibitemShut {NoStop}%
\bibitem [{\citenamefont {Stre{\v c}ka}\ \emph {et~al.}(2015)\citenamefont
  {Stre{\v c}ka}, \citenamefont {Kar{\v l}ov\'a},\ and\ \citenamefont
  {Madaras}}]{Strecka15}%
  \BibitemOpen
  \bibfield  {author} {\bibinfo {author} {\bibfnamefont {J.}~\bibnamefont
  {Stre{\v c}ka}}, \bibinfo {author} {\bibfnamefont {K.}~\bibnamefont {Kar{\v
  l}ov\'a}}, \ and\ \bibinfo {author} {\bibfnamefont {T.}~\bibnamefont
  {Madaras}},\ }\href {\doibase 10.1016/j.physb.2015.03.031} {\bibfield
  {journal} {\bibinfo  {journal} {Physica B}\ }\textbf {\bibinfo {volume}
  {466}},\ \bibinfo {pages} {76} (\bibinfo {year} {2015})}\BibitemShut
  {NoStop}%
\bibitem [{\citenamefont {Kar{\v l}ov\'a}\ \emph {et~al.}(2016)\citenamefont
  {Kar{\v l}ov\'a}, \citenamefont {Stre{\v c}ka},\ and\ \citenamefont
  {Madaras}}]{Karlova16}%
  \BibitemOpen
  \bibfield  {author} {\bibinfo {author} {\bibfnamefont {K.}~\bibnamefont
  {Kar{\v l}ov\'a}}, \bibinfo {author} {\bibfnamefont {J.}~\bibnamefont
  {Stre{\v c}ka}}, \ and\ \bibinfo {author} {\bibfnamefont {T.}~\bibnamefont
  {Madaras}},\ }\href {\doibase 10.1016/j.physb.2016.01.033} {\bibfield
  {journal} {\bibinfo  {journal} {Physica B}\ }\textbf {\bibinfo {volume}
  {488}},\ \bibinfo {pages} {49} (\bibinfo {year} {2016})}\BibitemShut
  {NoStop}%
\bibitem [{\citenamefont {Kar{\v l}ov\'a}\ \emph
  {et~al.}(2017{\natexlab{a}})\citenamefont {Kar{\v l}ov\'a}, \citenamefont
  {Stre{\v c}ka},\ and\ \citenamefont {Richter}}]{Karlova16-1}%
  \BibitemOpen
  \bibfield  {author} {\bibinfo {author} {\bibfnamefont {K.}~\bibnamefont
  {Kar{\v l}ov\'a}}, \bibinfo {author} {\bibfnamefont {J.}~\bibnamefont
  {Stre{\v c}ka}}, \ and\ \bibinfo {author} {\bibfnamefont {J.}~\bibnamefont
  {Richter}},\ }\href {\doibase 10.1088/1361-648X/aa53ab} {\bibfield  {journal}
  {\bibinfo  {journal} {J. Phys.: Condens. Matter}\ }\textbf {\bibinfo {volume}
  {29}},\ \bibinfo {pages} {125802} (\bibinfo {year}
  {2017}{\natexlab{a}})}\BibitemShut {NoStop}%
\bibitem [{\citenamefont {Kar{\v l}ov\'a}\ \emph
  {et~al.}(2017{\natexlab{b}})\citenamefont {Kar{\v l}ov\'a}, \citenamefont
  {Stre{\v c}ka},\ and\ \citenamefont {Madaras}}]{Karlova16-2}%
  \BibitemOpen
  \bibfield  {author} {\bibinfo {author} {\bibfnamefont {K.}~\bibnamefont
  {Kar{\v l}ov\'a}}, \bibinfo {author} {\bibfnamefont {J.}~\bibnamefont
  {Stre{\v c}ka}}, \ and\ \bibinfo {author} {\bibfnamefont {T.}~\bibnamefont
  {Madaras}},\ }\href {\doibase 10.12693/APhysPolA.131.630} {\bibfield
  {journal} {\bibinfo  {journal} {Acta Phys. Polon. A}\ }\textbf {\bibinfo
  {volume} {131}},\ \bibinfo {pages} {630} (\bibinfo {year}
  {2017}{\natexlab{b}})}\BibitemShut {NoStop}%
\bibitem [{\citenamefont {Konstantinidis}(2009)}]{NPK09}%
  \BibitemOpen
  \bibfield  {author} {\bibinfo {author} {\bibfnamefont {N.~P.}\ \bibnamefont
  {Konstantinidis}},\ }\href {\doibase 10.1103/PhysRevB.80.134427} {\bibfield
  {journal} {\bibinfo  {journal} {Phys. Rev. B}\ }\textbf {\bibinfo {volume}
  {80}},\ \bibinfo {pages} {134427} (\bibinfo {year} {2009})}\BibitemShut
  {NoStop}%
\bibitem [{\citenamefont {Konstantinidis}(2017{\natexlab{a}})}]{NPK17}%
  \BibitemOpen
  \bibfield  {author} {\bibinfo {author} {\bibfnamefont {N.~P.}\ \bibnamefont
  {Konstantinidis}},\ }\href {\doibase 10.1088/1361-648X/aa6bd4} {\bibfield
  {journal} {\bibinfo  {journal} {J. Phys.: Condens. Matter}\ }\textbf
  {\bibinfo {volume} {29}},\ \bibinfo {pages} {215803} (\bibinfo {year}
  {2017}{\natexlab{a}})}\BibitemShut {NoStop}%
\bibitem [{\citenamefont {Konstantinidis}(2018)}]{NPK18}%
  \BibitemOpen
  \bibfield  {author} {\bibinfo {author} {\bibfnamefont {N.~P.}\ \bibnamefont
  {Konstantinidis}},\ }\href {\doibase 10.1016/j.jmmm.2017.09.020} {\bibfield
  {journal} {\bibinfo  {journal} {J. Magn. Magn. Mater.}\ }\textbf {\bibinfo
  {volume} {449}},\ \bibinfo {pages} {55} (\bibinfo {year} {2018})}\BibitemShut
  {NoStop}%
\bibitem [{\citenamefont {Schr{\"o}der}\ \emph {et~al.}(2005)\citenamefont
  {Schr{\"o}der}, \citenamefont {Schmidt}, \citenamefont {Schnack},\ and\
  \citenamefont {Luban}}]{Schroeder05}%
  \BibitemOpen
  \bibfield  {author} {\bibinfo {author} {\bibfnamefont {C.}~\bibnamefont
  {Schr{\"o}der}}, \bibinfo {author} {\bibfnamefont {H.-J.}\ \bibnamefont
  {Schmidt}}, \bibinfo {author} {\bibfnamefont {J.}~\bibnamefont {Schnack}}, \
  and\ \bibinfo {author} {\bibfnamefont {M.}~\bibnamefont {Luban}},\ }\href
  {\doibase 10.1103/PhysRevLett.94.207203} {\bibfield  {journal} {\bibinfo
  {journal} {Phys. Rev. Lett.}\ }\textbf {\bibinfo {volume} {94}},\ \bibinfo
  {pages} {207203} (\bibinfo {year} {2005})}\BibitemShut {NoStop}%
\bibitem [{\citenamefont {Konstantinidis}(2015{\natexlab{a}})}]{NPK15}%
  \BibitemOpen
  \bibfield  {author} {\bibinfo {author} {\bibfnamefont {N.~P.}\ \bibnamefont
  {Konstantinidis}},\ }\href {\doibase 10.1088/0953-8984/27/7/076001}
  {\bibfield  {journal} {\bibinfo  {journal} {J. Phys.: Condens. Matter}\
  }\textbf {\bibinfo {volume} {27}},\ \bibinfo {pages} {076001} (\bibinfo
  {year} {2015}{\natexlab{a}})}\BibitemShut {NoStop}%
\bibitem [{\citenamefont {Konstantinidis}(2022)}]{NPK23-2}%
  \BibitemOpen
  \bibfield  {author} {\bibinfo {author} {\bibfnamefont {N.~P.}\ \bibnamefont
  {Konstantinidis}},\ }\href@noop {} {\  (\bibinfo {year} {2022})},\ \Eprint
  {http://arxiv.org/abs/cond-mat/2207.11077} {cond-mat/2207.11077} \BibitemShut
  {NoStop}%
\bibitem [{\citenamefont {Konstantinidis}(2021)}]{NPK21}%
  \BibitemOpen
  \bibfield  {author} {\bibinfo {author} {\bibfnamefont {N.~P.}\ \bibnamefont
  {Konstantinidis}},\ }\href {\doibase 10.1088/1361-648X/ac0477} {\bibfield
  {journal} {\bibinfo  {journal} {J. Phys.: Condens. Matter}\ }\textbf
  {\bibinfo {volume} {33}},\ \bibinfo {pages} {325801} (\bibinfo {year}
  {2021})}\BibitemShut {NoStop}%
\bibitem [{\citenamefont {Schulenburg}\ \emph {et~al.}(2002)\citenamefont
  {Schulenburg}, \citenamefont {Honecker}, \citenamefont {Schnack},
  \citenamefont {Richter},\ and\ \citenamefont {Schmidt}}]{Schulenburg02}%
  \BibitemOpen
  \bibfield  {author} {\bibinfo {author} {\bibfnamefont {J.}~\bibnamefont
  {Schulenburg}}, \bibinfo {author} {\bibfnamefont {A.}~\bibnamefont
  {Honecker}}, \bibinfo {author} {\bibfnamefont {J.}~\bibnamefont {Schnack}},
  \bibinfo {author} {\bibfnamefont {J.}~\bibnamefont {Richter}}, \ and\
  \bibinfo {author} {\bibfnamefont {H.-J.}\ \bibnamefont {Schmidt}},\ }\href
  {\doibase 10.1103/PhysRevLett.88.167207} {\bibfield  {journal} {\bibinfo
  {journal} {Phys. Rev. Lett.}\ }\textbf {\bibinfo {volume} {88}},\ \bibinfo
  {pages} {167207} (\bibinfo {year} {2002})}\BibitemShut {NoStop}%
\bibitem [{\citenamefont {Richter}\ \emph {et~al.}(2004)\citenamefont
  {Richter}, \citenamefont {Schulenburg}, \citenamefont {Honecker},
  \citenamefont {Schnack},\ and\ \citenamefont {Schmidt}}]{Richter04}%
  \BibitemOpen
  \bibfield  {author} {\bibinfo {author} {\bibfnamefont {J.}~\bibnamefont
  {Richter}}, \bibinfo {author} {\bibfnamefont {J.}~\bibnamefont
  {Schulenburg}}, \bibinfo {author} {\bibfnamefont {A.}~\bibnamefont
  {Honecker}}, \bibinfo {author} {\bibfnamefont {J.}~\bibnamefont {Schnack}}, \
  and\ \bibinfo {author} {\bibfnamefont {H.-J.}\ \bibnamefont {Schmidt}},\
  }\href {\doibase https://doi.org/10.1088/0953-8984/16/11/029} {\bibfield
  {journal} {\bibinfo  {journal} {J. Phys.: Condens. Matt.}\ }\textbf {\bibinfo
  {volume} {16}},\ \bibinfo {pages} {779} (\bibinfo {year} {2004})}\BibitemShut
  {NoStop}%
\bibitem [{\citenamefont {Schnack}\ \emph {et~al.}(2006)\citenamefont
  {Schnack}, \citenamefont {Schmidt}, \citenamefont {Honecker}, \citenamefont
  {Schulenburg},\ and\ \citenamefont {Richter}}]{Schnack06}%
  \BibitemOpen
  \bibfield  {author} {\bibinfo {author} {\bibfnamefont {J.}~\bibnamefont
  {Schnack}}, \bibinfo {author} {\bibfnamefont {H.-J.}\ \bibnamefont
  {Schmidt}}, \bibinfo {author} {\bibfnamefont {A.}~\bibnamefont {Honecker}},
  \bibinfo {author} {\bibfnamefont {J.}~\bibnamefont {Schulenburg}}, \ and\
  \bibinfo {author} {\bibfnamefont {J.}~\bibnamefont {Richter}},\ }\href
  {\doibase https://doi.org/10.1088/1742-6596/51/1/007} {\bibfield  {journal}
  {\bibinfo  {journal} {J. Phys. Confer. Ser.}\ }\textbf {\bibinfo {volume}
  {51}},\ \bibinfo {pages} {43} (\bibinfo {year} {2006})}\BibitemShut {NoStop}%
\bibitem [{\citenamefont {Nakano}\ and\ \citenamefont
  {Sakai}(2013)}]{Nakano13}%
  \BibitemOpen
  \bibfield  {author} {\bibinfo {author} {\bibfnamefont {H.}~\bibnamefont
  {Nakano}}\ and\ \bibinfo {author} {\bibfnamefont {T.}~\bibnamefont {Sakai}},\
  }\href {\doibase 10.7566/JPSJ.82.083709} {\bibfield  {journal} {\bibinfo
  {journal} {J. Phys. Soc. Jpn.}\ }\textbf {\bibinfo {volume} {82}},\ \bibinfo
  {pages} {083709} (\bibinfo {year} {2013})}\BibitemShut {NoStop}%
\bibitem [{\citenamefont {Nakano}\ \emph
  {et~al.}(2014{\natexlab{a}})\citenamefont {Nakano}, \citenamefont {Isoda},\
  and\ \citenamefont {Sakai}}]{Nakano14}%
  \BibitemOpen
  \bibfield  {author} {\bibinfo {author} {\bibfnamefont {H.}~\bibnamefont
  {Nakano}}, \bibinfo {author} {\bibfnamefont {M.}~\bibnamefont {Isoda}}, \
  and\ \bibinfo {author} {\bibfnamefont {T.}~\bibnamefont {Sakai}},\ }\href
  {\doibase 10.7566/JPSJ.83.053702} {\bibfield  {journal} {\bibinfo  {journal}
  {J. Phys. Soc. Jpn.}\ }\textbf {\bibinfo {volume} {83}},\ \bibinfo {pages}
  {053702} (\bibinfo {year} {2014}{\natexlab{a}})}\BibitemShut {NoStop}%
\bibitem [{\citenamefont {Nakano}\ \emph
  {et~al.}(2014{\natexlab{b}})\citenamefont {Nakano}, \citenamefont {Sakai},\
  and\ \citenamefont {Hasegawa}}]{Nakano14-1}%
  \BibitemOpen
  \bibfield  {author} {\bibinfo {author} {\bibfnamefont {H.}~\bibnamefont
  {Nakano}}, \bibinfo {author} {\bibfnamefont {T.}~\bibnamefont {Sakai}}, \
  and\ \bibinfo {author} {\bibfnamefont {Y.}~\bibnamefont {Hasegawa}},\ }\href
  {\doibase 10.7566/JPSJ.83.084709} {\bibfield  {journal} {\bibinfo  {journal}
  {J. Phys. Soc. Jpn.}\ }\textbf {\bibinfo {volume} {83}},\ \bibinfo {pages}
  {084709} (\bibinfo {year} {2014}{\natexlab{b}})}\BibitemShut {NoStop}%
\bibitem [{\citenamefont {Furuchi}\ \emph {et~al.}(2021)\citenamefont
  {Furuchi}, \citenamefont {Nakano}, \citenamefont {Todoroki},\ and\
  \citenamefont {Sakai}}]{Furuchi21}%
  \BibitemOpen
  \bibfield  {author} {\bibinfo {author} {\bibfnamefont {R.}~\bibnamefont
  {Furuchi}}, \bibinfo {author} {\bibfnamefont {H.}~\bibnamefont {Nakano}},
  \bibinfo {author} {\bibfnamefont {N.}~\bibnamefont {Todoroki}}, \ and\
  \bibinfo {author} {\bibfnamefont {T.}~\bibnamefont {Sakai}},\ }\href
  {\doibase https://doi.org/10.1088/2399-6528/ac3f7a} {\bibfield  {journal}
  {\bibinfo  {journal} {J. Phys. Commun.}\ }\textbf {\bibinfo {volume} {5}},\
  \bibinfo {pages} {125008} (\bibinfo {year} {2021})}\BibitemShut {NoStop}%
\bibitem [{\citenamefont {Furuchi}\ \emph {et~al.}(2023)\citenamefont
  {Furuchi}, \citenamefont {Nakano},\ and\ \citenamefont {Sakai}}]{Furuchi23}%
  \BibitemOpen
  \bibfield  {author} {\bibinfo {author} {\bibfnamefont {R.}~\bibnamefont
  {Furuchi}}, \bibinfo {author} {\bibfnamefont {H.}~\bibnamefont {Nakano}}, \
  and\ \bibinfo {author} {\bibfnamefont {T.}~\bibnamefont {Sakai}},\ }\href
  {\doibase https://doi.org/10.7566/JPSCP.38.011167} {\bibfield  {journal}
  {\bibinfo  {journal} {JPS Conf. Proc.}\ }\textbf {\bibinfo {volume} {38}},\
  \bibinfo {pages} {011167} (\bibinfo {year} {2023})}\BibitemShut {NoStop}%
\bibitem [{\citenamefont {Konstantinidis}(2016)}]{NPK16-1}%
  \BibitemOpen
  \bibfield  {author} {\bibinfo {author} {\bibfnamefont {N.~P.}\ \bibnamefont
  {Konstantinidis}},\ }\href {\doibase 10.1088/0953-8984/28/45/456003}
  {\bibfield  {journal} {\bibinfo  {journal} {J. Phys.: Condens. Matter}\
  }\textbf {\bibinfo {volume} {28}},\ \bibinfo {pages} {456003} (\bibinfo
  {year} {2016})}\BibitemShut {NoStop}%
\bibitem [{\citenamefont {Konstantinidis}(2023{\natexlab{b}})}]{NPK23}%
  \BibitemOpen
  \bibfield  {author} {\bibinfo {author} {\bibfnamefont {N.~P.}\ \bibnamefont
  {Konstantinidis}},\ }\href {\doibase 10.21468/SciPostPhysCore.6.2.042}
  {\bibfield  {journal} {\bibinfo  {journal} {SciPost Phys. Core}\ }\textbf
  {\bibinfo {volume} {6}},\ \bibinfo {pages} {042} (\bibinfo {year}
  {2023}{\natexlab{b}})}\BibitemShut {NoStop}%
\bibitem [{Note1()}]{Note1}%
  \BibitemOpen
  \bibinfo {note} {The maximum number of discontinuities for the $\omega $
  values considered occur at 0.2462 and 0.2464, 8 of the magnetization and 3 of
  the susceptibility, and at 0.2465, 0.247, 0.2475, and 0.248, 7 of the
  magnetization and 4 of the susceptibility.}\BibitemShut {Stop}%
\bibitem [{\citenamefont {Landau}\ and\ \citenamefont
  {Binder}(1981)}]{Landau81}%
  \BibitemOpen
  \bibfield  {author} {\bibinfo {author} {\bibfnamefont {D.~P.}\ \bibnamefont
  {Landau}}\ and\ \bibinfo {author} {\bibfnamefont {K.}~\bibnamefont
  {Binder}},\ }\href {\doibase https://doi.org/10.1103/PhysRevB.24.1391}
  {\bibfield  {journal} {\bibinfo  {journal} {Phys. Rev. B}\ }\textbf {\bibinfo
  {volume} {24}},\ \bibinfo {pages} {1391} (\bibinfo {year}
  {1981})}\BibitemShut {NoStop}%
\bibitem [{\citenamefont {Holtschneider}\ \emph {et~al.}(2007)\citenamefont
  {Holtschneider}, \citenamefont {Wessel},\ and\ \citenamefont
  {Selke}}]{Holtschneider07}%
  \BibitemOpen
  \bibfield  {author} {\bibinfo {author} {\bibfnamefont {M.}~\bibnamefont
  {Holtschneider}}, \bibinfo {author} {\bibfnamefont {S.}~\bibnamefont
  {Wessel}}, \ and\ \bibinfo {author} {\bibfnamefont {W.}~\bibnamefont
  {Selke}},\ }\href {\doibase https://doi.org/10.1103/PhysRevB.75.224417}
  {\bibfield  {journal} {\bibinfo  {journal} {Phys. Rev. B}\ }\textbf {\bibinfo
  {volume} {75}},\ \bibinfo {pages} {224417} (\bibinfo {year}
  {2007})}\BibitemShut {NoStop}%
\bibitem [{\citenamefont {Konstantinidis}(2017{\natexlab{b}})}]{NPK17-1}%
  \BibitemOpen
  \bibfield  {author} {\bibinfo {author} {\bibfnamefont {N.~P.}\ \bibnamefont
  {Konstantinidis}},\ }\href {\doibase 10.1088/2399-6528/aa95db} {\bibfield
  {journal} {\bibinfo  {journal} {J. Phys.: Comm.}\ }\textbf {\bibinfo {volume}
  {1}},\ \bibinfo {pages} {055027} (\bibinfo {year}
  {2017}{\natexlab{b}})}\BibitemShut {NoStop}%
\bibitem [{\citenamefont {Konstantinidis}(2015{\natexlab{b}})}]{NPK15-1}%
  \BibitemOpen
  \bibfield  {author} {\bibinfo {author} {\bibfnamefont {N.~P.}\ \bibnamefont
  {Konstantinidis}},\ }\href {\doibase 10.1140/epjb/e2015-60119-1} {\bibfield
  {journal} {\bibinfo  {journal} {Eur. Phys. J. B}\ }\textbf {\bibinfo {volume}
  {88}},\ \bibinfo {pages} {167} (\bibinfo {year}
  {2015}{\natexlab{b}})}\BibitemShut {NoStop}%
\bibitem [{\citenamefont {Konstantinidis}\ and\ \citenamefont
  {Coffey}(2001)}]{NPK01}%
  \BibitemOpen
  \bibfield  {author} {\bibinfo {author} {\bibfnamefont {N.~P.}\ \bibnamefont
  {Konstantinidis}}\ and\ \bibinfo {author} {\bibfnamefont {D.}~\bibnamefont
  {Coffey}},\ }\href {\doibase 10.1103/PhysRevB.63.184436} {\bibfield
  {journal} {\bibinfo  {journal} {Phys. Rev. B}\ }\textbf {\bibinfo {volume}
  {63}},\ \bibinfo {pages} {184436} (\bibinfo {year} {2001})}\BibitemShut
  {NoStop}%
\end{thebibliography}%

%\begin{table}
%\begin{center}
%\caption{Magnetization discontinuities of the AXXM (\ref{eqn:model}). The columns give the value of the magnetic field $h$ for which the discontinuity appears over its saturation value $h_{sat}$, the reduced magnetization $\frac{M}{N}$ below and above the discontinuity, and the reduced magnetization change. The saturation magnetic field $h_{sat}=3(3+\sqrt{5})$.}
%\begin{tabular}{c|c|c|c}
%$\frac{h}{h_{sat}}$ & $({\frac{M}{N}})_{below}$ & $({\frac{M}{N}})_{above}$ & $\frac{\Delta M}{N} (\times 10^{-3})$ \\
%\hline
% 0.0284380 & 0.0292860 & 0.0307879 & 1.50188 \\
%\hline
% 0.1108123 & 0.1187076 & 0.1211165 & 2.408849 \\
%\hline
% 0.1433832 & 0.1557494 & 0.1569121 & 1.16263488 \\
% \hline
% 0.1974717 & 0.2119367 & 0.2153885 & 3.45184 \\
%\hline
% 0.5034437 & 0.4977158 & 0.4995004 & 1.78462 \\
%\hline
% 0.5364572 & 0.5311187 & 0.5858516 & 54.732929
%\end{tabular}
%\label{table:classicalmagndiscAXXM}
%\end{center}
%\end{table}

\end{document}